\documentclass[prb,twocolumn,showpacs,amsmath,amssymb,superscriptaddress]{revtex4-1}
\usepackage{graphicx}
\usepackage{dcolumn}
\usepackage{bm}
\usepackage{bbm}
\usepackage{ulem}
\usepackage{color}

\begin{document}

\title{Topological Nonsymmorphic Crystalline Superconductors}

\author{Qing-Ze Wang}
\affiliation{Department of Physics, The Pennsylvania State University, University Park, Pennsylvania 16802-6300, USA}
\author{Chao-Xing Liu}
\affiliation{Department of Physics, The Pennsylvania State University, University Park, Pennsylvania 16802-6300, USA}

\date{\today}

\begin{abstract}
Topological superconductors possess a nodeless superconducting gap in the bulk and gapless zero energy modes, known as ``Majorana zero modes'', at the boundary of a finite system. In this work, we introduce a new class of topological superconductors, which are protected by nonsymmorphic crystalline symmetry and thus dubbed ``topological nonsymmorphic crystalline superconductors''. We construct an explicit Bogoliubov-de Gennes type of model for this superconducting phase in the D class and show how Majorana zero modes in this model are protected by glide plane symmetry. Furthermore, we generalize the classification of topological nonsymmorphic crystalline superconductors to the classes with time reversal symmetry, including the DIII and BDI classes, in two dimensions. Our theory provides a guidance to search for new topological superconducting materials with nonsymmorphic crystal structures.
\end{abstract}

\maketitle

\section{Introduction}
The research on topological superconductors (TSCs) has attracted intensive interests due to its gapless boundary excitations, known as the ``Majorana zero modes''\cite{ryu2002,fu2008,roy2008,qi2009,wilczek2009,law2009,lutchyn2010,hasan2010,qi2011,alicea2012,leijnse2012,mourik2012,rokhinson2012,das2012,beenakker2013}, with intrinsically non-local nature and exotic exchange statistics, and aims in the potential applications in low-decoherence quantum information processing and topological quantum computations\cite{ivanov2001,moore1991,nayak2008,alicea2011}. The search for new topological superconducting phases and materials is a substantial step for this goal.

The first classification of TSCs (and also other topological insulating phases) was achieved by Schnyder \textit{et al.}\cite{schnyder2008} based on Altland-Zirnbauer symmetry class \cite{zirnbauer1996,altland1997} for the systems with or without particle-hole symmetry (PHS), time reversal symmetry(TRS) and their combination, the so-called chiral symmetry. Later, it was realized that when additional symmetry exists in a system, new topological phases can be obtained, and the gapless edge/surface modes require the protection from additional symmetry. In particular, it has been shown that new topological insulating and superconducting phases emerge when the system has mirror symmetry and are dubbed ``topological mirror insulators''\cite{teo2008,hsieh2012,shiozaki2014,wang2015,sun2015} and ``topological mirror superconductors'' \cite{zhang2013,ueno2013,chiu2013,shiozaki2014}, respectively. Recent work has also revealed that nonsymmorphic symmetry, including glide plane symmetry and screw axis symmetry, can lead to new topological insulating phases, as well as topological semi-metal phases\cite{parameswaran2013,liu2014,fang2015,shiozaki2015,fang2015a}. In this work, we are interested in the role of the nonsymmorphic crystalline symmetry, mainly glide plane symmetry, in the classification of TSCs. We focus on the following three questions: (1) are there any topologically non-trivial phases that are protected by glide plane symmetry? (2) What's the difference between glide plane symmetry and mirror symmetry in the classification of TSCs? (3) What's the relationship between this superconducting phase and other TSCs? Below, we will first discuss the role of glide plane symmetry in the classification of superconducting gap functions, which indicates the possibility of topological superconductors protected by glide plane symmetry, thus dubbed ``topological nonsymmorphic crystalline superconductors (TNCSc)''. We also construct an explicit tight-binding model in the D class with boundary Majorana zero modes and demonstrate that the existence of Majorana zero modes comes from nonsymmorphic symmetry of this model. Finally, we discuss the relationship between TNSCs and weak TSCs and generalize TNCSc to the classes DIII and BDI with time reversal symmetry for both spinless and spin-$\frac{1}{2}$ fermions.

\section{nonsymmorphic symmetry and superconducting gap function}
In this section, we will first consider the role of glide plane symmetry in the classification of superconducting gap functions. We start from a generic Bogoliubov-de Gennes (BdG) type of Hamiltonian of superconductors with nonsymmorphic symmetry in the normal states, which can be written in the momentum space as
\begin{eqnarray}
\nonumber H
= \frac{1}{2}\sum_{\bold{k}}(c^{\dag}_{\bold{k}}, c^T_{-\bold{k}})  H_{BdG} \left(\begin{array}{c} c_{\bold{k}}\\c^{\dag T}_{-\bold{k}} \end{array}\right )
\label{ham}
\end{eqnarray}
with
\begin{eqnarray}
H_{BdG} = \left (\begin{array}{cc}
		h(\bold{k}) - \mu&\Delta(\bold{k})\\
		\Delta^{\dag}(\bold{k})&-h^*(-\bold{k})+\mu
\end{array}\right),
\label{hbdg}
\end{eqnarray}
where $h({\bf k})$ is for single-particle Hamiltonian of normal states, $\mu$ is the chemical potential and $\Delta$ denotes the superconducting gap function. $c_{\bold{k}}$ is an annihilation operator with $n$ components and we also use $c_{\bold{k},\alpha}$ ($\alpha = 1,...,n$) to denote each component with $\alpha = \{s,l\}$ for spins $s$ and orbitals(lattice sites) $l$. The superconducting gap function is related to annihilation operators by $\Delta_{\alpha,\beta}(\bold{k}) = V_0\langle c_{\bold{k},\beta} c_{-\bold{k},\alpha}\rangle$, where $V_0$ is the strength of attractive interactions. The BdG Hamiltonian satisfies the PHS $C H_{BdG}(\bold{k}) C^{-1} = - H_{BdG}(-\bold{k})$ with the PHS operator $C = \tau_1 \times \mathbb{I} K$, where $\tau_1$ is the first Pauli matrix acting on the Nambu space, $\mathbb{I}$ is an  $n\times n$ unit matrix and K is complex conjugation. The PHS (or Fermi statistics) requires the constraint $\Delta(\bold{k}) = - \Delta^T(-\bold{k})$ for the gap function.

Next, we consider how the nonsymmorphic symmetry yields constraint on the forms of single-particle Hamiltonian and superconducting gap functions in a nonsymmorphic crystal. Here we consider the glide plane symmetry, represented by $g = \{m|\bold{\tau}\}$ where $m$ is a mirror operator and $\bold{\tau}$ is a non-primitive translation operator along a direction within the mirror plane. For single-particle Hamiltonian, the glide plane symmetry requires
$D^\dag_{\bold{k}}(g) h(\bold{k}) D_{\bold{k}}(g) = h(g\bold{k})$, where $D_{\bold{k}}(g)$ is the representation matrix for glide plane symmetry at the momentum ${\bf k}$ and defined as $g c^{\dag}_{\bold{k},\alpha} g^{-1} = \sum_{\beta} D^*_{\bold{k},\alpha\beta}(g) c^{\dag}_{g\bold{k},\beta}$\cite{appendix}. Here we emphasize that the representation matrix for glide plane symmetry takes the form $D_{\bold{k}}(g) = e^{i\bold{k}\cdot \bold{\tau}}D(m)$, where $e^{i\bold{k}\cdot \bold{\tau}}$ is a phase factor due to a non-primitive translation and $D(m)$ is the projective representation of mirror operator $m$. For the case with only glide plane symmetry, all the projective representations are one dimensional (1D) and equivalent to the conventional representations.

The symmetry of the gap function $\Delta({\bf k})$ is determined by the Cooper pair wave functions, which transform as the direct product of the representation $D^\dag_{\bold{k}}(g)\otimes D^*_{-\bold{k}}(g)$\cite{appendix}. For the case with only glide plane symmetry, all the 1D representations can be labeled by $D_{\bold{k}}(g) = e^{i\bold{k}\cdot \bold{\tau}}  D(m) = \delta e^{i\bold{k}\cdot \bold{\tau}}$ where $\delta = \pm i$ for spin-$\frac{1}{2}$ systems and $\delta = \pm 1$ for spinless systems. Thus, the gap function should transform as $D^\dag_{\bold{k}}(g) \Delta(\bold{k}) D^*_{-\bold{k}}(g) = \eta \Delta(g\bold{k})$, where $\eta=\pm$ applies for both the spin-$\frac{1}{2}$ and spinless systems and depends on the nature of superconducting gap functions\cite{appendix}. We will show how to classify different superconducting gap functions based on glide plane symmetry explicitly for a model Hamiltonian in the next section.

We emphasize that the superconducting gap function may preserve ($\eta=+$) or spontaneously break ($\eta=-$) glide plane symmetry. Nevertheless, similar to the case of inversion symmetry\cite{fu2010,yang2014} or mirror symmetry\cite{ueno2013}, one can always re-define a glide plane symmetry operation as $G_{\eta}(\bold{k}) = Diag [D_{\bold{k}}(g), \eta D^*_{-\bold{k}}(g)]$ for the BdG type of Hamiltonian, which satisfies the condition $G^{-1}_{\eta}(\bold{k}) H_{BdG}(\bold{k}) G_{\eta}(\bold{k}) = H_{BdG}(g \bold{k})$. In this way, we can regard the BdG Hamiltonian as a semiconductor Hamiltonian with additional PHS.

\begin{figure}[tb]
	\includegraphics[width = 0.9\columnwidth,angle=0]{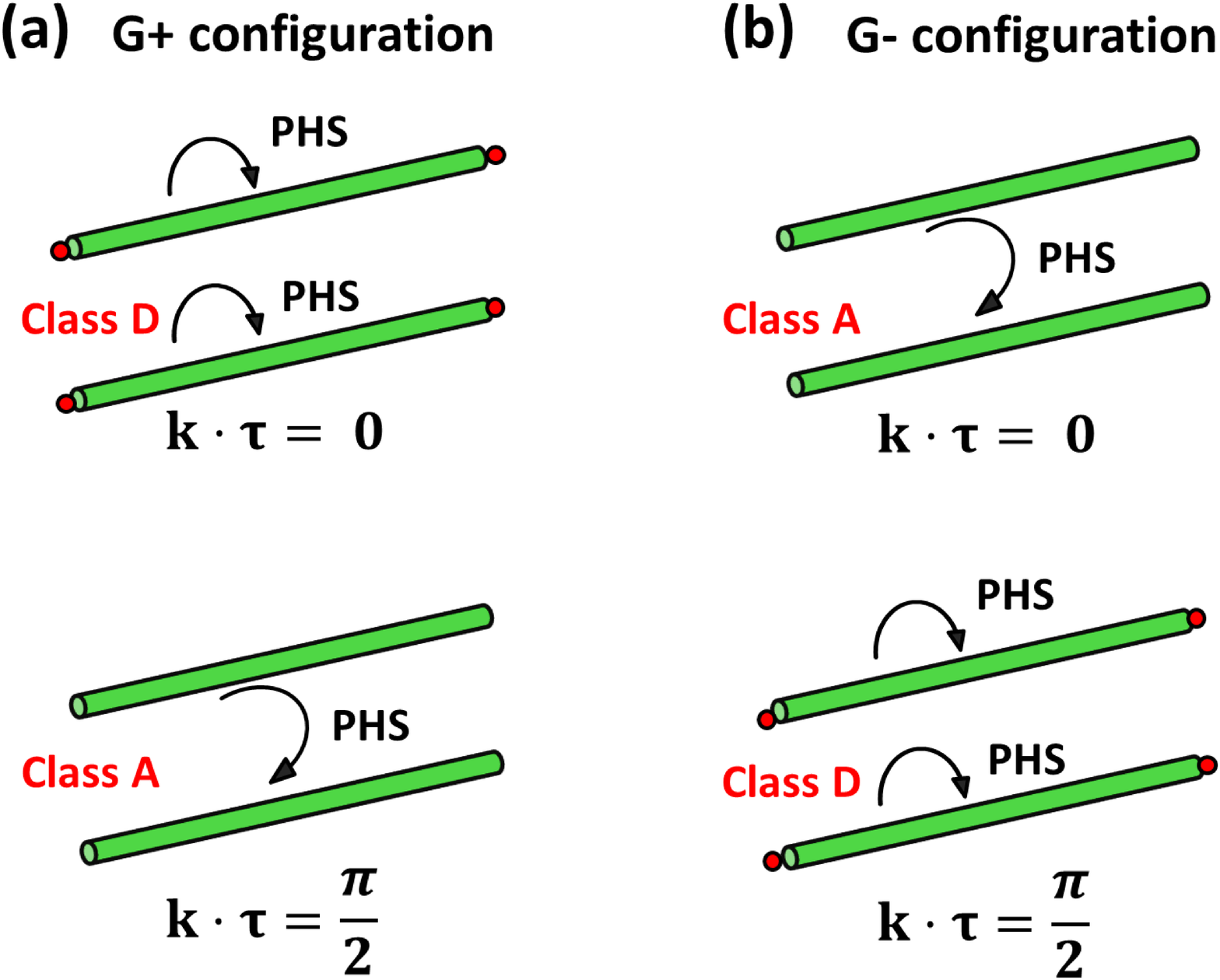}
  \caption{  (Color online). Two different configurations for $G_{\pm}(\bold{k})$. For $G_+$($G_-$), the Hamiltonian symmetry class is D(A) along $\bold{k} \cdot \bold{\tau} = 0$ lines; while along $\bold{k} \cdot \bold{\tau} = \frac{\pi}{2}$ lines, the Hamiltonian symmetry class is A(D). The red dots denote possible Majarona zero modes at ends of the lines.
  }
    \label{fig1}
\end{figure}

Due to the existence of the glide plane symmetry $G_\eta({\bf k})$, the eigenstates $\psi({\bf k})$ of the BdG Hamiltonian, $H_{BdG}\psi({\bf k})=E\psi({\bf k})$, can also be chosen to be the eigenstate of $G_\eta({\bf k})$, $G_{\eta}({\bf k})\psi({\bf k})=\delta_{\eta}e^{i\bold{k} \cdot \bold{\tau}}\psi({\bf k})$, on the glide invariant plane (GIP) in the momentum space, $g\bold{k} = \bold{k}$ (mod $\bold{P}$), where $\bold{P}$ is a reciprocal lattice vector. Here $\delta_{\eta}$ is given by $\pm$ ($\pm i$) for the spinless (spin-$\frac{1}{2}$) systems and we call the eigenvalue $\delta_{\eta} e^{i\bold{k} \cdot \bold{\tau}}$ as glide parity. Next, we look at the relationship of glide parities between one eigenstate $\psi({\bf k})$ and its partner $\tilde{\psi}(-\bold{k}) = C \psi(\bold{k})$ under PHS. Direct calculation gives $G_{\eta}(-\bold{k})\tilde{\psi}(-\bold{k}) =\eta\delta^*_{\eta}  e^{-i \bold{k} \cdot \bold{\tau}}\tilde{\psi}(-\bold{k})$ by using that $CG_{\eta}(\bold{k})C^{-1} = \eta G_{\eta}(-\bold{k})$\cite{appendix}. Therefore, $\psi_{\bold{k}}$ and its particle-hole partner $\tilde{\psi}_{-\bold{k}}$ possess glide parity $\delta_{\eta}  e^{i \bold{k} \cdot \bold{\tau}}$ and $\eta\delta^*_{\eta}  e^{-i \bold{k} \cdot \bold{\tau}}$, respectively. This leads to the conclusion as depicted in Fig. \ref{fig1}. When the gap function satisfies $G_+({\bf k})$ symmetry, for the spinless (spin-$\frac{1}{2}$) systems, $\psi_{\bold{k}}$ and its particle-hole partner $\tilde{\psi}_{-\bold{k}}$ share the same glide parity along the momentum line $\bold{k} \cdot \bold{\tau} = 0$ ($\bold{k} \cdot \bold{\tau} = \frac{\pi}{2}$) on the GIP, while they have opposite glide parities along the momentum line $\bold{k} \cdot \bold{\tau} = \frac{\pi}{2}$ ($\bold{k} \cdot \bold{\tau} = 0$) on the GIP. When the gap function satisfies $G_-({\bf k})$ symmetry, we find an opposite behavior for the momentum lines $\bold{k} \cdot \bold{\tau} = 0$ and $\bold{k} \cdot \bold{\tau} = \frac{\pi}{2}$, compared to the case of $G_+({\bf k})$ symmetry. We notice that the momentum line $\bold{k} \cdot \bold{\tau} = \frac{\pi}{2}$ corresponds to the BZ boundary since $2{\bf \tau}$ is a primitive lattice vector of the system.

Here we emphasize different roles of glide plane symmetry and mirror symmetry for the BdG Hamiltonian of superconductivity. For the glide plane symmetry $g = \{ m| \bold{\tau} \}$ and the corresponding mirror symmetry $m$, the GIP and the mirror invariant plane are the same. As shown in Ref. \onlinecite{ueno2013}, the PHS either preserves the subspace with a fixed mirror parity or transforms the subspace with one mirror parity to the other. In contrast, due to the additional phase factor from the non-primitive translation of glide plane symmetry, the behaviors of PHS acting on the glide parity subspaces are always opposite for the momentum lines $\bold{k} \cdot \bold{\tau} = 0$  and $\bold{k} \cdot \bold{\tau} = \frac{\pi}{2}$. This prevents us to define a topological invariant on the whole 2D GIP since two glide parity subspaces are always ``connected'' to each other. However, if we limit the glide parity subspace only on the momentum line $\bold{k} \cdot \bold{\tau} = 0$ or $\bold{k} \cdot \bold{\tau} = \frac{\pi}{2}$, the PHS will either preserve the glide parity subspace or transform the subspace with one glide parity to the other, similar to the case of mirror symmetry. This immediately suggests the possibility of defining topological invariants on the 1D momentum lines $\bold{k} \cdot \bold{\tau} = 0$ or $\bold{k} \cdot \bold{\tau} = \frac{\pi}{2}$ for superconductors with glide plane symmetry. Below, we will present explicitly a BdG type of model Hamiltonian with glide plane symmetry and show the existence of Majorana zero modes at the boundary. Then we will discuss bulk topological invariants and the corresponding topological classification.

\section{Model Hamiltonian in the D class}
\begin{figure}[tb]
	\includegraphics[width = 0.9\columnwidth,angle=0]{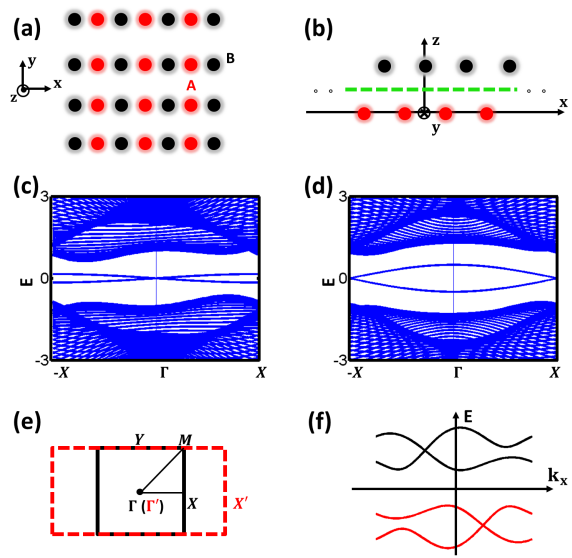}
  \caption{  (Color online). (a) and (b), Schematic plots of the lattice structure from top view and side view. They are 1D chains along y direction. There are two inequivalent atom sites, denoted as A(Red ball) and B(Black ball), respectively. A plane passing through the dashed green line is the glide plane.  (c) Edge modes for $G_{+}$ configuration with $\Delta_+$. (d) Edge modes for $G_{-}$ configuration with $\Delta_-$. (e) Brillouin zone(Black square) and extended Brillouin zone(Red dashed rectangle) defined by glide plane symmetry. (f) A general dispersion for a 1D chain with glide plane symmetry.
  }
    \label{fig2}
\end{figure}

Our spinless fermion model with glide plane symmetry is based on a two dimensional (2D) rectangle lattice with two sets of equivalent sites, as shown by A and B sites in Fig. \ref{fig2}(a) and (b). The glide plane symmetry operator is given by $g_z = \{m_z | \bold{\tau} = (\frac{a}{2},0, 0)\}$ with a reflection $m_z$ along the z direction followed by a translation of $a/2$ along the x direction ($a$ is a lattice constant), and relates the A sites to the B sites. The normal state Hamiltonian reads
\begin{eqnarray}
	\nonumber h({\bf k}) = && \epsilon(\bold{k}) \sigma_0 + t_3 cos(\frac{(k_x - \phi)a}{2})cos(\frac{k_xa}{2}) \sigma_1  \\
&& + t_3 cos(\frac{(k_x - \phi)a}{2})sin(\frac{k_xa}{2}) \sigma_2
\label{he}
\end{eqnarray}
on the basis $|A,\bold{k}\rangle$ and $|B,\bold{k}\rangle$, where $\epsilon(\bold{k}) = m_0 + t_1 cos(k_xa) + t_2 cos(k_ya)$, $\sigma_{0}$ is a 2$\times$2 unit matrix, $\sigma_i$ with $i = 1$, $2$, $3$ are Pauli matrices that describe the A and B sites and $\phi$ depends on the choice of orbitals \cite{appendix}. Furthermore, the glide plane symmetry operator on such a basis is $D_{\bold{k}}(g) = e^{i\frac{k_xa}{2}}(cos(\frac{k_xa}{2}) \sigma_1 + sin(\frac{k_xa}{2}) \sigma_2)$. One can easily check that $D_{\bold{k}}^2(g) = e^{ik_xa}$ and $D^{-1}_{\bold{k}}(g) H(k_x,k_y) D_{\bold{k}}(g) = H(k_x,k_y)$.

As discussed above, the gap functions can be classified according to glide plane symmetry and when the glide plane symmetry for the BdG Hamiltonian is $G_{\eta}$, the gap function satisfies three conditions: $\Delta^T(\bold{k}) = -\Delta(-\bold{k})$ (PHS); $D^\dag_{\bold{k}}(g) \Delta(\bold{k}) D^*_{-\bold{k}}(g) = \eta \Delta(g\bold{k})$ (glide plane symmetry) and $\Delta(\bold{k}) = \Delta(\bold{k} + \bold{G}) $\cite{appendix}. The complete classificaiton of gap functions for this model Hamiltonian is discussed in the Supplemental Material \cite{appendix}. Here we only consider two typical gap functions $\Delta_{+} = \Delta_0 sin(k_ya) \sigma_0$ and $\Delta_{-} = \Delta_0 sin(k_ya) \sigma_3$ with the symmetries $G_+$ and $G_-$, respectively. We take the BdG Hamiltonian (Eq. \ref{hbdg}) with the single-particle Hamiltonian (Eq. \ref{he}) and the gap function $\Delta_\pm$ and calculate energy dispersion of this Hamiltonian on a slab configuration. The slab is chosen to be infinite along the x direction and finite along the y direction, so that the glide plane symmetry $g_z$ is still preserved. The energy dispersion is shown in Fig. \ref{fig2} (c) for $\Delta_+$ and (d) for $\Delta_-$. In both cases, one can find two edge bands appearing in the bulk superconducting gap at one edge. However, these two edge bands cross at zero energy and give rise to Majorana zero modes at $\Gamma$ for $\Delta_+$ (Fig. \ref{fig2}(c)), but at $X$ for $\Delta_-$ (Fig. \ref{fig2}(d)).

The underlying physical reason of different positions of Majorana zero modes for these two cases comes from the relation between glide plane symmetry and PHS discussed in the last section. Let's take the case of the gap function $\Delta_+$ with $G_+$ symmetry as an example. The state $\psi_{\bf k}$ and its particle-hole partner $\tilde{\psi}_{-\bf k}$ share the same glide parity at $\Gamma$ ($k_x=0$), and thus it is possible for them to be the same state. Since PHS changes the energy $E$ of $\psi_{\bf k}$ to $-E$ of $\tilde{\psi}_{-\bf k}$, the eigen energy must be zero once they are the same state. This analysis also suggests that two Majorana zero modes at $\Gamma$ must belong to different glide parity subspace, and thus no coupling is allowed between them to open a gap. In contrast, the glide parities for $\psi_{\bf k}$ and $\tilde{\psi}_{-\bf k}$ are opposite at $X$ ($k_x=\frac{\pi}{a}$). Thus, these two states must be different at $X$ and PHS can not require their energies to be zero. This analysis can also be applied to $\Delta_-$ with $G_-$ symmetry and leads to the opposite conclusion. Another intuitive picture to prove non-trivial properties of 1D edge modes in Fig. \ref{fig2} (c) and (d) is to consider a general one dimensional superconductor with glide plane symmetry. As shown in Ref. \onlinecite{liu2014,fang2015,shiozaki2015}, due to the glide plane symmetry, all the bands must appear in pairs, as shown schematically by two black lines (two bands with opposite glide parities) in Fig. \ref{fig2} (f). Furthermore, the PHS of superconductivity requires two additional hole bands at the negative energy, as shown by two red lines in Fig. \ref{fig2} (f). Therefore, there must be even number of pairs of bands for a 1D nonsymmorphic superconductor. A single pair of bands shown in Fig. \ref{fig2} (c) and (d) can only exist at the 1D boundary of a 2D system. This gives the ``no-go'' theorem for nonsymmorphic superconductors\cite{wu2006}.

\section{Bulk topological invariants and the extended Brillouin zone}
The above analysis has shown that two momentum lines ${\bf k\cdot\tau}=0$ and ${\bf k\cdot\tau}=\frac{\pi}{2}$ play the essential role in the classification of TSCs in nonsymmorphic crystals. We can view the bulk Hamiltonian on ${\bf k\cdot\tau}=0$ or ${\bf k\cdot\tau}=\frac{\pi}{2}$ as a 1D Hamiltonian. For the case of $\Delta_+$ with the $G_+$ symmetry, the Hamiltonian $H_{BdG}$ (Eq. \ref{hbdg}) has PHS along the line ${\bf k\cdot\tau}=0$ for each glide parity subspace, thus belonging to the D class, while it has no PHS along the line ${\bf k\cdot\tau}=\frac{\pi}{2}$ for each glide parity subspace, as shown in Fig. \ref{fig1} (a). Since two glide parity subspaces are decoupled, one $\mathbb{Z}_2$ topological invariant of the D class can be defined on the line ${\bf k\cdot\tau}=0$ in the glide parity subspace for a 1D Hamiltonian. In contrast, for the case of $\Delta_-$ with the $G_-$ symmetry, one $\mathbb{Z}_2$ topological invariant can be defined on the line ${\bf k\cdot\tau}=\frac{\pi}{2}$. In our example, we can re-write the BdG Hamiltonian with the eigenstates of $G_\pm$ as a basis and one can see immediately for the case with the $G_+$ ($G_-$) symmetry, the Hamiltonian is exactly equivalent to the 1D Kitaev model of p-wave superconductors\cite{kitaev2001} in each glide parity subspace when $k_x=0$ ($k_x=\frac{\pi}{a}$)\cite{appendix}.

More insights about this system can be obtained from the view of the extended Brillouin zone (BZ)\cite{lee2008}, which has been widely used in the field of iron pnictide superconductors. For nonsymmorphic crystals, all the eigenstates of the Hamiltonian can be labeled by the eigenvalues of glide operators, defined as $G_{\eta}\psi(\tilde{\bf k})=e^{i\tilde{\bf k}\cdot \tau}\psi(\tilde{\bf k})$, in which $\tilde{\bf k}$ is called ``pseudocrystal momentum''\cite{lee2008} and defines the extended BZ. For our model, the glide plane symmetry operation only involves translation by $\frac{a}{2}$ along the x direction, and thus the extended BZ for $\tilde{k}_x$ is doubled along the x direction ($\tilde{k}_x \in [-\frac{2\pi}{a},\frac{2\pi}{a}]$), compared to the conventional BZ for $k_x$, as shown in Fig. \ref{fig2} (e). Since we have $G_{\eta}\psi({\bf k})=\delta_{\eta}e^{i{\bf k\cdot \tau}}\psi({\bf k})$, this suggests that the pseudocrystal momentum $\tilde{\bf k}$ is related to momentum ${\bf k}$ by $\tilde{\bf k}={\bf k}$ when $\delta_{\eta}=+$ and $\tilde{\bf k}={\bf k}+{\bf Q}$ with ${\bf Q\cdot\tau}=\pm\pi$ when $\delta_{\eta}=-$. Here the sign of ${\bf Q\cdot\tau}$ is determined by keeping $\tilde{k}_x$ in the region $[-\frac{2\pi}{a},\frac{2\pi}{a}]$ and $k_y$ in the region $[-\frac{\pi}{a},\frac{\pi}{a}]$. As a result, the BdG Hamiltonian can also be rewritten as $H^{ex}_{BdG}(\tilde{\bf k})=H_{BdG,+}({\bf k})=H_{BdG,+}(\tilde{\bf k})$ for $\tilde{k}_x\in [-\frac{\pi}{a},\frac{\pi}{a}]$ and $H^{ex}_{BdG}(\tilde{\bf k})=H_{BdG,-}({\bf k})=H_{BdG,-}(\tilde{\bf k}-{\bf Q})$ for $\tilde{k}_x \in [\frac{\pi}{a},\frac{2\pi}{a}]$ and $\tilde{k}_x\in [-\frac{2\pi}{a},-\frac{\pi}{a}]$ in the extended BZ. Here $H_{BdG,\pm}$ is the BdG Hamiltonian in the subspace with glide parity $\pm e^{i{\bf k\cdot\tau}}$. For our model Hamiltonian, $H_{BdG,\pm}$ corresponds to the two by two Hamiltonian defined in the Supplemental Material\cite{appendix}. For the case of $\Delta_+$, the form of $H^{ex}_{BdG}$ is given by $H^{ex}_{BdG}(\tilde{\bold{k}}) = -(\epsilon(\tilde{\bold{k}}) - \mu + t_3 cos(\frac{\tilde{k}_xa}{2})cos(\frac{\phi a}{2})) \tau_3 + t_3 sin(\frac{\tilde{k}_xa}{2})sin(\frac{\phi a}{2}) \tau_0 + \Delta_0 sin(k_ya)\tau_1$, where $\epsilon(\tilde{\bold{k}}) = m_0 + t_1 cos(\tilde{k}_xa) + t_2 cos(k_ya)$. We notice that if we take the hopping parameters $t_1$ and $t_3$ along the x direction to be zero, this Hamiltonian exactly corresponds to the 1D Kitaev chain with one Majorana zero mode at the open boundary\cite{kitaev2001}. With the hopping along the x direction, all the 1D Kitaev chains are coupled along the x direction, so Majorana zero modes at the end of the chains couple to each other and expand into a band. This corresponds to the weak TSCs\cite{teo2010,hughes2014}, which is in analogy to weak topological insulators\cite{fu2007}. The PHS requires $E(\tilde{\bf k})=-E(-\tilde{\bf k})$ for the band of Majorana zero modes. Therefore, zero energy states can only appear for $\tilde{k}_x=0$ and $\tilde{k}_x=\frac{2\pi}{a}$ ($\tilde{k}_x$ is periodic in $\frac{4\pi}{a}$), which both correspond to $k_x=0$ in the conventional BZ. In contrast, for the case of $\Delta_-$, the gap function comes from the so-called $\eta$ pairing for two electrons with the momenta $\tilde{\bf k}$ and ${\bf Q}-\tilde{\bf k}$ to form a Cooper pair\cite{yang1989,hu2012,hu2013,hao2014,wang2015g,appendix}
(${\bf Q}=(\frac{2\pi}{a},0)$ for our model). In this case, the PHS requires $E(\tilde{\bf k})=-E({\bf Q}-\tilde{\bf k})$ for the Majorana band, leading to the zero energy states at $\tilde{k}_x=\pm \frac{\pi}{a}$. This analysis based on the extended BZ is consistent with our previous results and show explicitly the relationship between TNSCs and weak TSCs.

\section{Discussion and conclusion}
The above results for TNCSc can be directly generalized to the systems with spin-$\frac{1}{2}$ and with additional time reversal (TR) symmetry. For spin-$\frac{1}{2}$ systems, since $\delta_\eta$ in the glide parity is given by $\pm i$, there is an additional minus sign when considering how the glide parity of an eigenstate of the BdG Hamiltonian transforms under PHS. This leads to the consequence that the $\mathbb{Z}_2$ topological invariant can be defined at $\bold{k}\cdot \bold{\tau} = \frac{\pi}{2}$ ($\bold{k}\cdot \bold{\tau} = 0$) for the systems with the $G_+$ ($G_-$) symmetry. According to the standard topological classification, TR symmetry can change the symmetry class from the $D$ class to $BDI$ for spinless systems and $DIII$ for spin-$\frac{1}{2}$ systems. To see how it affects the classification of TNCSc, we consider an example of a spin-$\frac{1}{2}$ system in the $DIII$ class with the $G_+$ symmetry. If we take a state $\psi(\bold{k})$ with glide parity $\delta e^{i\bold{k}\cdot \bold{\tau}}$ where $\delta = i$, the glide parity of its PHS partner has been shown to be $\delta^* e^{-i\bold{k}\cdot \bold{\tau}}$ and the glide parity of its TR partner is also $\delta^* e^{-i\bold{k}\cdot \bold{\tau}}$, where $\Theta G_+(\bold{k}) \Theta^{-1} = G_+(-\bold{k})$ is used and TR operator is $\Theta = \left ( \begin{array}{cc} \Theta_e & 0 \\ 0&\Theta^{\dag T}_e \end{array} \right)$ with $\Theta_e = i\sigma_0s_2K$ and $s_2$ the second Pauli matrix acting on spin space. One can see that chiral symmetry $\Pi = C \times \Theta$ exists in each glide parity subspace for any momentum.  In addition, at the momentum line ${\bf k\cdot \tau}=\frac{\pi}{2}$, PHS and TRS also exist in each glide parity subspace. Therefore, the symmetry class is DIII for the momentum line ${\bf k\cdot \tau}=\frac{\pi}{2}$ and AIII for other momentum lines (${\bf k\cdot \tau} \neq \frac{\pi}{2}$) in each glide parity subspace. This leads to $\mathbb{Z}_2$$\bold{\oplus \mathbb{Z}_2}$ classification at ${\bf k\cdot \tau}=\frac{\pi}{2}$, $\mathbb{Z}$ classification at ${\bf k\cdot \tau}=0$ and $\mathbb{Z}\oplus \mathbb{Z}$ classification at other momentum lines for the whole BdG Hamiltonian\cite{schnyder2008,appendix}, in sharp contrast to the $\mathbb{Z} \times \mathbb{Z}$ classification of topological mirror superconductors in the DIII class\cite{zhang2013}. This classification leads to the existence of edge flat bands in the DIII class for TNCSc (See Supplemental materials \cite{appendix}). The spinless and spin-$\frac{1}{2}$ TNCSc in classes D, DIII and BDI are also studied in the Supplemental Material\cite{appendix}.
Nonsymmorphic symmetry is known to exist in several classes of superconducting materials, including iron pnictide superconductors\cite{kamihara2008,stewart2011,dagotto2013,cvetkovic2013,qing2012,he2013,tan2013}, BiS$_2$-based layered superconductors\cite{mizuguchi2012,mizuguchi2012a,usui2012,yazici2013,jha2013,xing2012,yang2013,lin2013,dai2015}, and heavy fermion superconductors\cite{stewart1984,stewart2011}, \textit{e.g.} UPt$_3$\cite{joynt2002}, UBe$_{13}$\cite{ott1983}. Our topological classification of TNCSc can be directly applied to these systems to search for realistic topological superconducting materials.

\begin{acknowledgments}
We would like to thank X. Dai, X.Y. Dong, Ken Shiozaki, Fan Zhang and Jiangping Hu for the helpful discussions.
\end{acknowledgments}

{\it Note added. - } After finishing this paper, we notice a paper on arxiv\cite{hao2015}, which concerns possible topological superconducting phases in monolayer FeSe and potential relation to nonsymmorphic symmetry. We also notice another recent paper on arxiv\cite{shiozaki20152} about topological classification of TNCSc based on the twisted equivariant K-theory.

\begin{appendix}
\begin{widetext}

\section{Bogoliubov-de Gennes Hamiltonian and the glide plane symmetry}
We start from a generic mean-field Hamiltonian of superconductors with nonsymmorphic symmetry in the normal states, which can be written in the momentum space as
\begin{eqnarray}
\nonumber H && = \sum_{\bold{k},\alpha,\beta} c^\dag_{\bold{k},\alpha} (h_{\alpha,\beta}(\bold{k})- \mu) c_{\bold{k},\beta} + \sum_{\bold{k},\alpha,\beta}\frac{1}{2}(\Delta^{\dag}_{\alpha,\beta} c_{-\bold{k},\alpha}c_{\bold{k},\beta} + \Delta_{\alpha,\beta}c^{\dag}_{\bold{k},\alpha}c^\dag_{-\bold{k},\beta})\\
&&= \frac{1}{2}\sum_{\bold{k}}(c^{\dag}_{\bold{k}}, c^T_{-\bold{k}})  H_{BdG} \left(\begin{array}{c} c_{\bold{k}}\\c^{\dag T}_{-\bold{k}} \end{array}\right )
\label{ham}
\end{eqnarray}
with
\begin{eqnarray}
H_{BdG} = \left (\begin{array}{cc}
		h(\bold{k}) - \mu&\Delta(\bold{k})\\
		\Delta^{\dag}(\bold{k})&-h^*(-\bold{k})+\mu
\end{array}\right),
\label{ahbdg}
\end{eqnarray}
where $h({\bf k})$ is for single-particle Hamiltonian of normal states, $\mu$ is the chemical potential and $\Delta$ denotes the superconducting gap function. $c_{\bold{k}}$ is an annihilation operator with $n$ components and we also use $c_{\bold{k},\alpha}$ ($\alpha = 1,...,n$) to denote each component with $\alpha = \{s,l\}$ for spins $s$ and orbitals(lattice sites) $l$. The superconducting gap function is related to annihilation operators by $\Delta_{\alpha,\beta}(\bold{k}) = V_0\langle c_{\bold{k},\beta} c_{-\bold{k},\alpha}\rangle$, where $V_0$ is the strength of attractive interactions. The Bogoliubov-de gennes (BdG) Hamiltonian satisfies the particle-hole symmetry (PHS) $C H_{BdG}(\bold{k}) C^{-1} = - H_{BdG}(-\bold{k})$ with the PHS operator $C = \tau_1 \times \mathbb{I} K$, where $\tau_1$ is Pauli matrix acting on the Nambu space, $\mathbb{I}$ is an  $n\times n$ unit matrix and K is complex conjugation. The PHS (or Fermi statistics) requires the constraint $\Delta(\bold{k}) = - \Delta^T(-\bold{k})$ for the gap function.

glide plane symmetry can be expressed as $g = \{m|\bold{\tau}\}$ with $m$ a mirror operator and $\bold{\tau}$ a non-primitive translation operator along the direction within the mirror plane. We have $g c^{\dag}_{\bold{k},\alpha} g^{-1} = \sum_{\beta} D^*_{\bold{k},\alpha\beta}(g) c^{\dag}_{g\bold{k},\beta}$ and $g c_{\bold{k},\alpha} g^{-1} = \sum_{\beta} D_{\bold{k},\alpha\beta}(g) c_{g\bold{k},\beta}$ where $D_{\bold{k}}(g)$ is representation of space group $g$ in the little group of wave-vector $\bold{k}$. Further, $D_{\bold{k}}(\{ m|\bold{\tau}\}) = e^{i\bold{k}\cdot \bold{\tau}}D(m)$, where $D_m$ is the projective representation of mirror operator $m$.

Since the normal state $H_{e} = \sum_{\alpha\beta} c^\dag_{\bold{k},\alpha} h_{\alpha\beta}(\bold{k}) c_{\bold{k},\beta}$ has the glide plane symmetry, $g H_{e} g^{-1} = H_{e}$. Explicitly,
\begin{eqnarray}
\nonumber \sum_{\bold{k},\alpha\beta} g c^{\dag}_{\bold{k},\alpha} g^{-1} h_{\alpha\beta}(\bold{k}) g c_{\bold{k},\alpha} g^{-1} &&= \sum_{\bold{k},\alpha,\beta} D^*_{\bold{k},\alpha\gamma}(g) c^{\dag}_{g\bold{k},\gamma} h_{\alpha\beta}(\bold{k})D_{\bold{k},\beta\lambda}(g) c_{g\bold{k},\lambda} \\
\nonumber&&=  \sum_{\bold{k},\gamma\lambda} c^{\dag}_{\bold{k},\gamma} (\sum_{\alpha\beta}D^*_{g^{-1}\bold{k},\alpha\gamma}(g) h_{\alpha\beta}(g^{-1}\bold{k})D_{g^{-1}\bold{k},\beta\lambda}(g)) c_{\bold{k},\lambda}
\end{eqnarray}
Therefore, we have
\begin{eqnarray}
\sum_{\alpha\beta}D^*_{g^{-1}\bold{k},\alpha\gamma}(g) h_{\alpha\beta}(g^{-1}\bold{k})D_{g^{-1}\bold{k},\beta\lambda}(g) = h_{\gamma,\lambda}(\bold{k})
\end{eqnarray}
More elegantly, the normal state Hamiltonian under glide plane symmetry satisfies
\begin{eqnarray}
D^\dag_{\bold{k}}(g) h(\bold{k}) D_{\bold{k}}(g) = h(g\bold{k}).
\end{eqnarray}
We also have $D^\dag_{\bold{k}}(g) = D^{-1}_{\bold{k}}(g)$.

The pairing terms in Eq. \ref{ham} fulfill the requirement of glide plane symmetry. However, the gap functions are not necessary to respect the glide plane symmetry.
\begin{eqnarray}
\nonumber g (\sum_{\bold{k},\alpha\beta} \Delta_{\alpha\beta}(\bold{k}) c^\dag_{\bold{k},\alpha}c^\dag_{-\bold{k},\beta})g^{-1} = \sum_{\bold{k},\alpha\beta\gamma\lambda} D^*_{\bold{k},\alpha\gamma}(g)D^*_{-\bold{k},\beta\lambda}(g) \Delta_{\alpha\beta}(\bold{k})c^\dag_{g\bold{k},\alpha}c^\dag_{-g\bold{k},\beta} = \sum_{\bold{k},\gamma\lambda} \tilde{\Delta}_{\gamma\lambda}(\bold{k})c^\dag_{\bold{k},\gamma}c^\dag_{-\bold{k},\lambda}
\end{eqnarray}
, where $\tilde{\Delta}_{\gamma\lambda}(\bold{k})$ is the transformed gap function. Thus, we arrive at $\tilde{\Delta}_{\gamma\lambda}(\bold{k})  = \sum_{\alpha\beta} D^*_{g^{-1}\bold{k},\alpha\gamma}(g)D^*_{-g^{-1}\bold{k},\beta\lambda}(g) \Delta_{\alpha\beta}(g^{-1}\bold{k})$. This indicates that the gap function $\Delta(\bold{k})$ transforms according to the decomposition of the direct product of the representation $D^\dag_{\bold{k}}(g)\otimes D^*_{-\bold{k}}(g)$.

If we choose the basis as the eigenstates of glide plane symmetry operator, all representations are reduced to one dimension(1D) and $D_{\bold{k}}(g) = e^{i\bold{k}\cdot \bold{\tau}}  D(m) = e^{i\bold{k}\cdot \bold{\tau}} \delta$ where $\delta = \pm i$ for spin-$\frac{1}{2}$ systems and  $\delta = \pm 1$ for spinless systems. Thus, $\tilde{\Delta}_{\gamma\lambda}(\bold{k})  = D^*_{g^{-1}\bold{k},\gamma\gamma}(g)D^*_{-g^{-1}\bold{k},\lambda\lambda}(g) \Delta_{\alpha\beta}(g^{-1}\bold{k}) = e^{-ig^{-1}\bold{k}\cdot \bold{\tau}}\delta^*_{\gamma}e^{ig^{-1}\bold{k}\cdot \bold{\tau}}\delta^*_{\lambda}\Delta_{\alpha\beta}(g^{-1}\bold{k}) = \delta^*_{\gamma}\delta^*_{\lambda} \Delta_{\alpha\beta}(g^{-1}\bold{k})$. We can write down the requirement of gap function under nonsymmorphic symmetry in a compact way
\begin{eqnarray}
\tilde{\Delta}(g\bold{k}) = D^\dag_{\bold{k}}(g) \Delta(\bold{k}) D^*_{-\bold{k}}(g) = \eta \Delta(g\bold{k})
\label{delta}
\end{eqnarray}
where $\eta = \pm 1$ applies for both spin-$\frac{1}{2}$ and spinless systems.

The matrix form of the BdG Hamiltonian is shown in Eq. \ref{ahbdg} in the basis of Nambu space $\Psi(\bold{k}) = \left( \begin{array} {c} c_{\bold{k}}\\ c^{\dag T}_{-\bold{k}} \end{array} \right)$. According to Eq. \ref{delta}, we can treat the BdG Hamiltonian as a semiconductor Hamiltonian with additional PHS and re-define the glide plane symmetry operator as
\begin{eqnarray}
G_{\eta}(\bold{k}) = \left( \begin{array} {cc} D_{\bold{k}}(g)&0\\ 0& \eta D^*_{-\bold{k}}(g) \\ \end{array} \right)
\end{eqnarray}

Next we check how the BdG Hamiltonian transforms under glide plane symmetry by considering possible gap functions required in Eq. \ref{delta}. For the case with
$D^\dag_{\bold{k}}(g) \Delta(\bold{k}) D^*_{-\bold{k}}(g) = \eta \Delta(g\bold{k})$, we have
\begin{eqnarray}
\nonumber G^{-1}_{\eta}(\bold{k}) H_{BdG}(\bold{k}) G_{\eta}(\bold{k}) &&=  \left( \begin{array} {cc} D^{-1}_{\bold{k}}(g)&0\\ 0& \eta D^{*,-1}_{-\bold{k}}(g) \\ \end{array} \right) \left( \begin{array} {cc} h(\bold{k}) - \mu&\Delta(\bold{k})\\ \Delta(\bold{k})& -h^*(-\bold{k}) + \mu \\ \end{array} \right) \left( \begin{array} {cc} D_{\bold{k}}(g)&0\\ 0& \eta D^*_{-\bold{k}}(g) \\ \end{array} \right)\\
\nonumber && = \left( \begin{array} {cc} h(g\bold{k}) - \mu& D^{-1}_{\bold{k}}(g) \Delta(\bold{k}) \eta D^{*}_{-\bold{k}}(g)\\ h.c. & -h^*(-g\bold{k}) + \mu \\ \end{array} \right)\\
\nonumber && = \left( \begin{array} {cc} h(g\bold{k}) - \mu& \eta^2 \Delta(g\bold{k}) \\ h.c. & -h^*(-g\bold{k}) + \mu \\ \end{array} \right)\\
\nonumber && = \left( \begin{array} {cc} h(g\bold{k}) - \mu&\Delta(g\bold{k}) \\ h.c. & -h^*(-g\bold{k}) + \mu \\ \end{array} \right)\\
&& = H_{BdG}(g\bold{k})
\end{eqnarray}
where $D^\dag_{\bold{k}}(g) = D^{-1}_{\bold{k}}(g)$ is used. This gives us the form of symmetry transformation for the BdG Hamiltonian.

Due to the glide plane symmetry, all the eigenstates of the BdG Hamiltonian at the glide invariant plane (GIP) can be also expressed as the eigenstates of glide plane symmetry and the corresponding eigenvalues are dubbed ``glide parity'', as discussed in the main text. To show the glide parities of a state and its particle-hole partner, we take an example of the case with the $G_+$ symmetry. If the gap function preserves the glide plane symmetry, the BdG Hamiltonian commutes with $G_{+}(\bold{k})$ on GIPs. This indicates that one can simultaneously block diagonalize $H_{BdG}$ and $G_{+}(\bold{k})$ with a set of common eigenvectors. Each block owns a glide parity $\delta_+ e^{i\bold{k} \cdot \bold{\tau}}$ with $\delta_+ = \pm 1(\pm i)$ for spinless(spin-$\frac{1}{2}$) systems.

We start from that $CG_+(\bold{k}) = \left( \begin{array} {cc} 0&1\\1&0\\ \end{array} \right) K \left( \begin{array} {cc} D_{\bold{k}}(g)&0\\0&D^*_{-\bold{k}}(g)\\ \end{array} \right) = \left( \begin{array} {cc} D_{-\bold{k}}(g)&0\\0&D^*_{\bold{k}}(g)\\ \end{array} \right) \left( \begin{array} {cc} 0&1\\1&0\\ \end{array} \right) K = G_+(-\bold{k})$, i.e. $C G_{+}(\bold{k}) C^{-1} = G_{+}(-\bold{k})$, where C is the PHS operator. One can pick up a common eigenstate $\psi(\bold{k})$ of $H_{BdG}$ and $G_{+}(\bold{k})$ with glide parity $\delta_+ e^{i \bold{k} \cdot \bold{\tau}}$. Its particle-hole partner is denoted as $\tilde{\psi}(-\bold{k}) = C \psi(\bold{k})$. Then we have $G_{+}(-\bold{k})\tilde{\psi}(-\bold{k}) = G_{+}(-\bold{k})C \psi(\bold{k}) = C (G_{+}(\bold{k})\psi(\bold{k})) = C(\delta_+ e^{i \bold{k} \cdot \bold{\tau}} \psi(\bold{k})) = \delta^*_+  e^{-i \bold{k} \cdot \bold{\tau}}\tilde{\psi}(-\bold{k})$. Therefore, $\psi_{\bold{k}}$ and its PHS partner $\tilde{\psi}_{-\bold{k}}$ possess glide parity $\delta_+  e^{i \bold{k} \cdot \bold{\tau}}$ and $\delta^*_+  e^{-i \bold{k} \cdot \bold{\tau}}$, respectively.

\section{Model Hamiltonian}
In this section, we will show how to construct a tight-binding model for the TNCSc. We use $\phi(\bold{r}-\bold{R}_i-\bold{r}_{\alpha})$ to denote L$\ddot{o}$wding orbital $\alpha$ at $\bold{r}$, where $\bold{R}_i$ is the position of $i^{th}$ unit cell and $\bold{r}_\alpha$ is the position of atom site or orbital inside a unit cell. The Bloch wave function is defined as $\psi_{\bold{k},\alpha}(\bold{r}) = \frac{1}{\sqrt{N}} \sum_{\bold{R}_i} e^{i \bold{k}\cdot \bold{R}_i}\phi(\bold{r}-\bold{R}_i-\bold{r}_{\alpha})$ by using the linear combination of atomic orbitals(LCAO). It should be emphasized that the phase factor $e^{i \bold{k}\cdot \bold{R}_i}$ in our construction does not include the position ${\bf r}_{\alpha}$. This Bloch wave function can be written as $\psi_{\bold{k},\alpha}(\bold{r}) = e^{i\bold{k}\cdot \bold{r}} u_{\bold{k},\alpha}(\bold{r})$ where $u_{\bold{k},\alpha}(\bold{r}) = \frac{1}{\sqrt{N}} \sum_{\bold{R}_i} e^{i\bold{k}\cdot (\bold{R_i - r})} \phi(\bold{r}-\bold{R}_i-\bold{r}_{\alpha})$. It is easily checked that $u_{\bold{k},\alpha}(\bold{r} + \bold{R}) = \frac{1}{\sqrt{N}} \sum_{\bold{R}_i} e^{i\bold{k}\cdot (\bold{R_i - r - \bold{R}})} \phi(\bold{r}+ \bold{R}-\bold{R}_i-\bold{r}_{\alpha}) = \frac{1}{\sqrt{N}} \sum_{\bold{R}_i} e^{i\bold{k}\cdot ((\bold{R_i - \bold{R}) - r})} \phi(\bold{r}+ (\bold{R}_i-\bold{R})-\bold{r}_{\alpha}) =  \frac{1}{\sqrt{N}} \sum_{\bold{\delta R}} e^{i\bold{k}\cdot (\bold{\delta R - r})} \phi(\bold{r}+ \bold{\delta R}-\bold{r}_{\alpha}) =  u_{\bold{k},\alpha}(\bold{r})$. Thus, the Bloch theorem holds for such a choice of LCAO. Under such a choice of LCAO, we have that $\psi_{\bold{k},\alpha}(\bold{r}) = \psi_{\bold{k +P},\alpha}(\bold{r})$, where $\bold{P}$ is a reciprocal lattice vector. Any Hamiltonian on such a basis $H_{\alpha,\beta}(\bold{k}) = \int d\bold{r} \psi^*_{\bold{k},\alpha}(\bold{r}) \hat{H}\psi_{\bold{k},\beta}(\bold{r})$ satisfies that $H_{\alpha,\beta}(\bold{k+P}) = \int d\bold{r} \psi^*_{\bold{k+P},\alpha}(\bold{r}) \hat{H}\psi_{\bold{k+P},\beta}(\bold{r}) = \int d\bold{r} \psi^*_{\bold{k},\alpha}(\bold{r}) \hat{H}\psi_{\bold{k},\beta}(\bold{r}) = H_{\alpha,\beta}(\bold{k})$, i.e.
\begin{eqnarray}
H(\bold{k+P}) = H(\bold{k}).
\end{eqnarray}
Similarly, for the gap function, $\Delta(\bold{k+P}) = \langle \int d\bold{r} \psi_{\bold{k+P},\alpha}(\bold{r})\psi_{\bold{k+P},\alpha}(\bold{r})\rangle = \langle \int d\bold{r} \psi_{\bold{k},\alpha}(\bold{r})\psi_{\bold{k},\alpha}(\bold{r})\rangle = \Delta(\bold{k})$, i.e.
\begin{eqnarray}
\Delta(\bold{k+P}) = \Delta(\bold{k})
\end{eqnarray}.

Next let us take $c^\dag_{\bold{k},\alpha}$ and $c_{\bold{k},\alpha}$($c^\dag_{\bold{r},\alpha}$ and $c_{\bold{r},\alpha}$) to present creation and annihilation operators of $\psi_{\bold{k},\alpha}(\bold{r})$($\phi(\bold{r - R_i - r_\alpha})$) and show the real space form of the tight-binding model discussed in the section ``Model Hamiltonian'' of the main text. 
The normal state tight-binding Hamiltonian for the lattice structure reads
\begin{eqnarray}
H_e = &&m_0 \sum_{i,s = {A,B}} c^{\dag}_{i,s} c_{i,s} + [\sum_{i,s = {A,B}} (\frac{t_1}{2} c^{\dag}_{i+d_x,s} c_{i,s} +  \frac{t_2}{2} c^{\dag}_{i+d_y,s} c_{i,s}) + \sum_{i} \frac{t_3}{2}( c^{\dag}_{i,A} c_{i,B} + c^{\dag}_{i,A} c_{i-d_x,B}) +  H.c.]
\end{eqnarray}
where $i = \{i_x,i_y\}$ denotes index of unit cells, $d_{x,y}$ are primitive lattice vectors along x and y direction, A and B denote two inequivalent atom sites and H.c. represents their conjugation parts.  We further obtain a tight-binding model in the momentum space by performing an unusual Fourier transformation $c^\dag_{\bold{k},s} = \frac{1}{\sqrt{N}} \sum_i e^{i\bold{k}\cdot \bold{R_i}} c^\dag_{i,s}$ and $ c_{i,s} = \frac{1}{\sqrt{N}} \sum_i e^{i\bold{k}\cdot \bold{R_i}} c_{\bold{k},s}$, where $\bold{R}_i$ is the position of the i$^{th}$ unit cell. Such a Fourier transformation simplifies the Hamiltonian and leads to $H_e(\bold{k}+\bold{P}) = H_e(\bold{k})$ with $\bold{P}$ is a reciprocal lattice vector.

The gap functions for $G_{\pm}$ configurations mentioned previously need to satisfy three conditions: (1) PHS
\begin{eqnarray}
\Delta^T(\bold{k}) = -\Delta(-\bold{k});\label{appendix_PHS}
\end{eqnarray}
(2) glide plane symmetry
\begin{eqnarray}
D^\dag_{\bold{k}}(g) \Delta(\bold{k}) D^*_{-\bold{k}}(g) = \eta \Delta(g\bold{k});\label{appendix_glide}
\end{eqnarray}
and (3)
\begin{eqnarray}
\Delta(\bold{k}) = \Delta(\bold{k} + \bold{G}).\label{appendix_periodical}
 \end{eqnarray}
 These three conditions allow us to classify all the possible gap function for this model Hamiltonian. Let us define the gap functions $\Delta_i(\bold{k}) = f_i(\bold{k})\Gamma_i(\bold{k})$, where $i=1,...,4$, $f_i(\bold{k})$ is a complex function of $\bold{k}$ and $\Gamma_i(\bold{k})$ are four 2$\times$2 matrices, as shown in the second column in Table \ref{pair}.

The second and third column in Table \ref{pair} shows the ``parity'' of $\Gamma$ matrices in the sense of PHS and glide plane symmetry. Due to the PHS (\ref{appendix_PHS}), the parity of $f_i(\bold{k})$ is determined by the parity of $\Gamma$ from the second column, which is listed in the fourth column. For 2D system, $\bold{k} = (k_x,k_y)$ if $g = \{ m_z |\bold{\tau}\}$. For the last two columns, the existence of $f_i(k)$ is determined by Eq. (\ref{appendix_glide}) and $\xi_g$ in the third column. Let us take the $\Gamma|_1$ matrix with $\xi_g=1$ as an example. For $G_+$, since glide operation does not act on $f_1(k)$, we obtain $f_1(k)=f_1(k)$ and there is no constraint on $f_1$. But for $G_-$, Eq. (\ref{appendix_glide}) and $\xi_g=1$ together requires $f_1(k)=-f_1(k)$, leading to $f_1(k)=0$. Thus, no term is possible for $G_-$. Similar analysis can be applied to other matrices.
Based on the parity of $f_i(\bold{k})$ on the fourth column, we can get possible polynomials of $f_i(\bold{k})$, as listed in Table \ref{pairf}.

The parameters for the calculation of energy dispersion of the BdG Hamiltonian are shown in Table \ref{par}.

\begin{center}
\begin{table}[htb]
  \centering
  \begin{minipage}[t]{1\linewidth}
	  \caption{ Here $\xi_C$, $\xi_g$ and $\xi_f$ are defined by $\Gamma^T_i(\bold{k}) = \xi_C \Gamma_i(-\bold{k})$, $D^\dag_{\bold{k}}(g)\Gamma_i(\bold{k}) D^*_{-\bold{k}}(g) = \xi_g \Gamma_i(\bold{k})$, $f_i(\bold{k}) = \xi_f f_i(-\bold{k})$. PHS and glide plane symmetry requirements on $f_i(\bold{k})$ provide the $\xi_C = \pm 1$ and $\xi_g = \pm 1$ on the second and third columns.
The $\xi_f = \pm 1$ in the fourth column is the parity of $f_i(\bold{k})$ and obtained from the PHS (\ref{appendix_PHS}) and $\xi_C$.
NA in the fifth and sixth columns represents ``not available''.
Here $\Gamma_1 = \sigma_0$, $\Gamma_2 = \sigma_3$, $\Gamma_3 = cos(\frac{k_xa}{2})\sigma_1 + sin(\frac{k_xa}{2})\sigma_2$ and
$\Gamma_4 = sin(\frac{k_xa}{2})\sigma_1 - cos(\frac{k_xa}{2})\sigma_2$. }
\hspace{-1cm}
\begin{tabular}
[c]{cccccc}\hline\hline
                    & $\xi_C$ \vline& $\xi_g$ \vline& $\xi_f$ \vline& $G_+$: $f_i(\bold{k})$ \vline& $G_-$: $f_i(\bold{k})$\\\hline
$\Gamma_1$& + &+ & -& Valid &NA\\
$\Gamma_2$& + &- & -& NA &Valid\\
$\Gamma_3$&+ &+ & -& Valid&NA \\
$\Gamma_4$&- &- & +& NA&Valid \\\hline\hline
\label{pair}
\end{tabular}
  \end{minipage}
\end{table}
\end{center}

\begin{center}
\begin{table}[htb]
  \centering
  \begin{minipage}[t]{1.\linewidth}
	  \caption{possible polynomials of $f_i(\bold{k})$ for each $\Delta_i(\bold{k})$. NAs in the fourth and fifth columns represent ``not available''}
\hspace{-1cm}
\begin{tabular}
[c]{ccc}\hline\hline
                     & $G_+$: $f_i(\bold{k})$ & $G_-$: $f_i(\bold{k})$ \\\hline
$\Delta_1(\bold{k}) = f_1(\bold{k})\Gamma_1 $&$sin(k_xa)$, $sin(k_ya)$ &NA \\
$\Delta_2(\bold{k}) = f_2(\bold{k})\Gamma_2 $&NA & $sin(k_xa)$, $sin(k_ya)$\\
$\Delta_3(\bold{k}) = f_3(\bold{k})\Gamma_3 $&$sin(\frac{k_xa}{2})$ & NA \\
$\Delta_4(\bold{k}) = f_4(\bold{k})\Gamma_4 $&NA &$cos(\frac{k_xa}{2})$ \\\hline\hline
\label{pairf}
\end{tabular}
  \end{minipage}
\end{table}
\end{center}

\begin{center}
\begin{table}[htb]
  \centering
  \begin{minipage}[t]{1.\linewidth}
	  \caption{Parameters for the emergence of edge modes in Fig. 2(c) and (d) for $G_{\pm}$ configurations in the main text.}
\hspace{-1cm}
\begin{tabular}
[c]{cccccccc}\hline\hline
      &$m_0$& $t_1$& $t_2$& $t_3$ & $\mu$ & $\phi$ & $\Delta_0$\\\hline
$G_+$ & 1.5& -0.5& -3& -1& 0& 0.1$\pi$& 2\\
$G_-$ & 1.5& -0.5& -3& -0.5& 0& 0.1$\pi$& 2\\\hline\hline
\label{par}
\end{tabular}
  \end{minipage}
\end{table}
\end{center}

\section{Hamiltonian in the extended Brillouin zone for $G_{\pm}$}
In this section, we will analyze our model Hamiltonian in the extended Brillouin zone.
For the case with the $G_+$ symmetry, $G_+(\bold{k}) = e^{i\frac{k_xa}{2}} (cos(\frac{k_xa}{2})\tau_0 \otimes \sigma_1 + sin(\frac{k_xa}{2})\tau_0 \otimes \sigma_2)$. The eigenvalues are $\mp e^{i\frac{k_xa}{2}}$. The eigenvectors are $u_{1,-} = (0,0,-e^{-i\frac{k_xa}{2}},1)^T$, $u_{2,-} = (-e^{-i\frac{k_xa}{2}},1,0,0)^T$, $u_{1,+} = (0,0,e^{-i\frac{k_xa}{2}},1)^T$ and $u_{2,+} = (e^{-i\frac{k_xa}{2}},1,0,0)^T$, where $u_{\mp}$ corresponds to eigenvalues $\mp e^{i\frac{k_xa}{2}}$. On the above eigenvectors, the BdG Hamiltonian in the case with the $G_+$ symmetry can be expressed in a block diagonal matrix, which reads
\begin{eqnarray}
H_{BdG,g+} = \left (\begin{array}{cccc}
x_{-,g+}&\Delta_0sin(k_ya)&0&0\\
\Delta_0sin(k_ya)&y_{-,g+}&0&0\\
0&0&x_{+,g+}&\Delta_0sin(k_ya)\\
0&0&\Delta_0sin(k_ya)&y_{+,g+}\\
\end{array}\right)
\end{eqnarray}
where $x_{\pm,g+} = -(\epsilon(\bold{k}) - \mu) \mp t_3 cos(\frac{k_x +\phi}{2})$ and  $y_{\pm,g+} = (\epsilon(\bold{k}) - \mu) \pm t_3 cos(\frac{(k_x -\phi)a}{2})$. We can easily check that at $k_x = 0$, each block Hamiltonian owns PHS, which is the 1D Kitaev model for p-wave superconductors. Based on the above block diagonal Hamiltonian, one can see that it can be written as
\begin{eqnarray}
H^{ex}_{BdG}(G_+,\tilde{\bold{k}}) = \left( \begin{array}{cc}
x_{+,g+}(\tilde{\bold{k}})& \Delta_0sin(k_ya)\\
\Delta_0sin(k_ya) & y_{+,g+}(\tilde{\tilde{\bold{k}}})\\
\end{array} \right)
\end{eqnarray}
in the extended Brillouin zone, where $\tilde{k}_x \subset [-\frac{2\pi}{a},\frac{2\pi}{a}]$ and $k_y \subset [-\frac{\pi}{a},\frac{\pi}{a}]$. At momenta $\tilde{k}_x = 0, \frac{2\pi}{a}$, the Hamiltonian respects PHS. Thus, we can define a $\mathbb{Z}_2$ topological invariant $\nu_{g+}$, which is expressed as $(-1)^{\nu_{g+}} = sign(|t_2| - |\tilde{\mu}_+|)$\cite{kitaev2001,teo2010,chiu2013}, where $\tilde{\mu}_+ = \mu - m_0 -t_1-t_3cos(\frac{\phi a}{2})$ at $\tilde{k}_x = 0$ and $\tilde{\mu}_+ = \mu - m_0 -t_1+t_3cos(\frac{\phi a}{2})$ at $\tilde{k}_x = \pm \frac{2\pi}{a}$.  When we fold the extended BZ, $\tilde{k}_x =0,  \pm\frac{2\pi}{a}$ are all mapped at $k_x = 0$. Therefore, possible Majorana zero modes appear at $k_x = 0$. It should be pointed out that the topological invariants are different at $\tilde{k}_x = 0, \frac{2\pi}{a}$ by choosing appropriate parameters $t_3$ and $\phi$.

For the case with the $G_-$ symmetry, $G_-(\bold{k}) = e^{i\frac{k_xa}{2}} (cos(\frac{k_xa}{2})\tau_z \otimes \sigma_1 + sin(\frac{k_xa}{2})\tau_z \otimes \sigma_2)$. The eigenvalues are $\mp e^{i\frac{k_xa}{2}}$. The eigenvectors are $v_{1,-} = (0,0,e^{-i\frac{k_xa}{2}},1)^T$, $v_{2,-} = (-e^{-i\frac{k_xa}{2}},1,0,0)^T$, $v_{1,+} = (0,0,-e^{-i\frac{k_xa}{2}},1)^T$ and $v_{2,+} = (e^{-i\frac{k_xa}{2}},1,0,0)^T$, where $v_{\mp}$ corresponds to eigenvalues $\mp e^{i\frac{k_xa}{2}}$. On the above eigenvectors, our BdG Hamiltonian can be expressed in a block diagonal matrix, which reads
\begin{eqnarray}
H_{BdG,g-} = \left (\begin{array}{cccc}
x_{-,g-}&-\Delta_0sin(k_ya)&0&0\\
-\Delta_0sin(k_ya)&y_{-,g-}&0&0\\
0&0&x_{+,g-}&-\Delta_0sin(k_ya)\\
0&0&-\Delta_0sin(k_ya)&y_{+,g-}\\
\end{array}\right)
\end{eqnarray}
where $x_{\pm,g-} = -(\epsilon(\bold{k}) - \mu) \pm t_3 cos(\frac{(k_x +\phi)a}{2})$ and  $y_{\pm,g-} = (\epsilon(\bold{k}) - \mu) \pm t_3 cos(\frac{(k_x -\phi)a}{2})$. We can easily check that at $k_x = \frac{\pi}{a}$, each block Hamiltonian owns PHS, which is the 1D Kitaev model for p-wave superconductors.

The Hamiltonian for $G_-$ in the extended BZ can be written as
\begin{eqnarray}
H^{ex}_{BdG}(G_-,\tilde{\bold{k}}) = \left( \begin{array}{cc}
x_{+,g-}(\tilde{\bold{k}})& -\Delta_0sin(k_ya)\\
-\Delta_0sin(k_ya) & y_{+,g-}(\tilde{\tilde{\bold{k}}})\\
\end{array} \right)
\end{eqnarray}
where $\tilde{k}_x \subset [-\frac{2\pi}{a},\frac{2\pi}{a}]$ and $k_y \subset [-\frac{\pi}{a},\frac{\pi}{a}]$. At momentum $\tilde{k}_x = \pm \frac{\pi}{a}$, the Hamiltonian respects PHS. The  $\mathbb{Z}_2$ topological invariant $\nu_{g-}$ is defined as $(-1)^{\nu_{g-}} = sign(|t_2| - |\tilde{\mu}_+|)$, where $\tilde{\mu}_+ = \mu - m_0 -t_1\mp t_3sin(\frac{\phi a}{2})$ at momenta $\tilde{k}_x = \pm \frac{\pi}{a}$. When we fold the extended BZ, $\tilde{k}_x = \pm \frac{\pi}{a}$ are mapped at $k_x = \pm \frac{\pi}{a}$. Therefore, possible Majorana zero modes appear at $k_x = \pm \frac{\pi}{a}$.

Another feature that can be extracted from the extended BZ is that the pairing of Cooper pairs is between two electrons with the momenta ($\tilde{k}$,-$\tilde{k}$) for the case of the $G_+$ symmetry, making Majorana zero modes appear at $\tilde{k}_x = 0$, while the pairing is between two electrons with the momenta ($\tilde{k}$,$\bold{Q} - \tilde{k}$) with $\bold{Q} = (\frac{2\pi}{a},0)$, which is known as the $\eta$ pairing, for the case of the $G_-$ symmetry. For the $\eta$ pairing case, the Majorana zero modes emerge at $\tilde{k}_x = \frac{\bold{Q}}{2} = \frac{\pi}{a}$, where $\tilde{k}_x = -\tilde{k}_x + \bold{Q}$. We do the Fourier transformation of the gap functions from momentum space to real space for only x direction. For normal paring functions in the model of extended BZ, $\Delta(\delta r_x)_{norm} = \frac{1}{\sqrt{2\pi}} \sum_{\tilde{k}_x} \Delta(\tilde{k}_x,k_y) e^{i\tilde{k}_x \delta r_x} = \frac{1}{N} \sum_{i_x} \langle c_{i_x,k_y} c_{i_x - \delta r_x,-k_y}\rangle$, where $\Delta(\bold{k}) = \langle c_{\bold{k}}c_{-\bold{k}}\rangle$ and $c_{\bold{k}} = \frac{1}{\sqrt{N}} \sum_{i_x} c_{i_x,k_y} e^{-i\tilde{k}_xi_x}$ are used. This result indicates that the pairings along y direction are the same for A and B sites if $\delta r_x = 0$. On the other hand, for $\eta$ paring functions $\Delta(\delta r_x)_{\eta} = \frac{1}{\sqrt{2\pi}} \sum_{\tilde{k}_x} \Delta(\bold{Q} - \tilde{k}_x,k_y) e^{i\tilde{k}_x \delta r_x} = \frac{1}{N} \sum_{i_x} e^{-i \bold{Q} \cdot \bold{i_x}}\langle c_{i_x,k_y} c_{i_x - \delta r_x,-k_y}\rangle = \frac{1}{N} \sum_{i_x} e^{-i n\pi}\langle c_{i_x,k_y} c_{i_x - \delta r_x,-k_y}\rangle$, where integers $n = \{0, ..., 2N\}$ denote index of sites along x direction. Thus, the $\eta$ pairings along y direction are of opposite signs for A and B sites if $\delta r_x = 0$.

\section{Topological classification for two-dimensional superconductors with nonsymmorphic crystalline symmetry}

\subsection{Superconductors without time reversal symmetry}
The BdG Hamiltonian for both spin-$\frac{1}{2}$ and spinless superconductors without TRS belongs to symmetry class D.
Table \ref{class1} lists topological classifications for two-dimensional(2D) TNCSc (glide plane symmetry here) in the symmetry class D, in which the topological classification is  $\mathbb{Z}$ in 2D. When the glide plane symmetry is included, the topological invariant can be described by two 1D topological invariants at the momentum lines ${\bf k\cdot \tau}=0,\frac{\pi}{2}$. We list the corresponding symmetry classes and also their topological invariants along these momentum lines for $G_\eta$ in the last two rows of Table \ref{class1}. In the following subsections, we analyze the Hamiltonian symmetry classes at momentum lines ${\bf k\cdot \tau}=0, \frac{\pi}{2}$ to obtain topological classification\cite{schnyder2008}. In this section, we focus on the case with a zero Chern number in the whole 2D Brillouin zone and the $\mathbb{Z}_2$ topological invariants in the two glide parity subspaces would be the same, leading to only one $\mathbb{Z}_2$ invariant at momentum lines ${\bf k\cdot \tau}=0, \frac{\pi}{2}$. The case with non-zero Chern number in the whole 2D Brillouin zone is discussed in Sec. V(D) of the Supplemental Material and also in Ref. \onlinecite{shiozaki20152}.


\begin{center}
\begin{table}[htb]
\centering
\begin{minipage}[t]{1\linewidth}
\caption{topological classification of 2D nonsymmorphic crystalline superconductors(glide plane symmetry here) without TRS. The class of Hamiltonian and its corresponding topological classification are listed in one block. $NoGS$ in the first column represents a system has no glide plane symmetry. `-' stands for no topological classification.}
\hspace{-2cm}
\label{class1}
\begin{tabular}
[c]{|c|c|c|c|c|}\hline
&\multicolumn{2}{|c|}{spinless} &\multicolumn{2}{|c|}{spin-$\frac{1}{2}$}\\\hline
NoGS(2D)&\multicolumn{2}{|c|}{D, $\mathbb{Z}$} & \multicolumn{2}{|c|}{D, $\mathbb{Z}$} \\\hline
& $\bold{k} \cdot \bold{\tau} = 0$ & $\bold{k} \cdot \bold{\tau} = \frac{\pi}{2}$& $\bold{k} \cdot \bold{\tau} = 0$ & $\bold{k} \cdot \bold{\tau} = \frac{\pi}{2}$\\\hline
$G_+$ & D, $\mathbb{Z}_2$   &A, -               & A, -              & D, $\mathbb{Z}_2$\\\hline
$G_-$ & A, -                &D, $\mathbb{Z}_2$ & D, $\mathbb{Z}_2$ & A, - \\ \hline
\end{tabular}
  \end{minipage}
\end{table}
\end{center}

\subsubsection{class D: $G_+$ for spinless superconductors}
For $C G_+(\bold{k}) C^{-1} = G_+(-\bold{k})$, we select a state $\psi_{\bold{k}}$ with glide parity $\delta e^{i \bold{k} \cdot \bold{\tau}}$ with $\delta = \pm 1$. Its PH partner $\tilde{\psi}_{-\bold{k}}$ has glide parity $\delta^* e^{-i \bold{k} \cdot \bold{\tau}}$. Thus, at the momentum line $\bold{k} \cdot \bold{\tau} = 0$, PHS exists in each glide parity subspace. Therefore, the symmetry class in each glide parity subspace is D and topological invariant in 1D is $\mathbb{Z}_2$. At $\bold{k} \cdot \bold{\tau} = \frac{\pi}{2}$, PHS interchanges states in opposite glide parity subspaces. Thus, the symmetry class in each glide parity subspace is A and there is no topological classification in 1D.

\subsubsection{class D: $G_-$ for spinless superconductors}
For $C G_-(\bold{k}) C^{-1} = -G_-(-\bold{k})$, we consider a state $\psi_{\bold{k}}$ with glide parity $\delta e^{i \bold{k} \cdot \bold{\tau}}$ with $\delta = \pm 1$. Its PH partner $\tilde{\psi}_{-\bold{k}}$ has glide parity $-\delta^* e^{-i \bold{k} \cdot \bold{\tau}}$. Thus, at $\bold{k} \cdot \bold{\tau} = 0$, PHS interchanges states in two glide parity subspaces and the symmetry class in each glide parity subspace is A without topological classification in 1D. At $\bold{k} \cdot \bold{\tau} = \frac{\pi}{2}$, there is PHS in each glide parity subspace, leading to a $\mathbb{Z}_2$ classification in each parity subspace.

\subsubsection{class D: $G_+$ for spin-$\frac{1}{2}$ superconductors}
For $C G_+(\bold{k}) C^{-1} = G_+(-\bold{k})$, the PH partner $\tilde{\psi}_{-\bold{k}}$ of a state $\psi_{\bold{k}}$ with glide parity $\delta e^{i \bold{k} \cdot \bold{\tau}}$ ($\delta = \pm i$) has glide parity $\delta^* e^{-i \bold{k} \cdot \bold{\tau}}$. Thus, at $\bold{k} \cdot \bold{\tau} = 0$, PHS interchanges states in two glide parity subspaces, giving rise to symmetry class A without topological classification in 1D in each glide parity subspace. At $\bold{k} \cdot \bold{\tau} = \frac{\pi}{2}$, PHS exists in each glide parity subspace, leading to the symmetry class D and $\mathbb{Z}_2$ classification in 1D in each glide parity subspace.

\subsubsection{class D: $G_-$ for spin-$\frac{1}{2}$ superconductors}
For $C G_-(\bold{k}) C^{-1} = -G_-(-\bold{k})$, the PHS partner $\tilde{\psi}_{-\bold{k}}$ of $\psi_{\bold{k}}$ with glide parity $\delta e^{i \bold{k} \cdot \bold{\tau}}$ ($\delta = \pm i$) has glide parity $-\delta^* e^{-i \bold{k} \cdot \bold{\tau}}$. Thus, at $\bold{k} \cdot \bold{\tau} = 0$, PHS exists in each glide parity subspace, leading to the symmetry class D in each glide parity subspace and the corresponding $\mathbb{Z}_2$ classification in 1D. At $\bold{k} \cdot \bold{\tau} = \frac{\pi}{2}$, similar analysis suggests symmetry class A and no topological classification in each glide parity subspace.

\subsection{Superconductors with time reversal symmetry}
In this section, we consider the BdG Hamiltonian for spinless(spin-$\frac{1}{2}$) superconductors with TRS belonging to symmetry class BDI(DIII). We emphasize that tiem reversal symmetry always commutes with any space group symmetry in a physical system. Table \ref{class2} lists possible topological classifications for 2D nonsymmorphic crystalline superconductors(glide plane symmetry here) in the symmetry classes BDI and DIII. The topological classification of 2D superconductors for class DIII is  $\mathbb{Z}_2$ while there is no topological classification for the class BDI in 2D. In our case, for $G_-$, we only need to concern the two momentum lines $\bold{k\cdot \tau} = 0, \frac{\pi}{2}$, similar to the case of D symmetry class, while for $G_+$, we find the chiral symmetry exists for any momentum. This makes the classification of the $G_+$ case quite different from other cases. Below we will discuss each case separately.


\begin{center}
\begin{table}[htb]
\centering
\begin{minipage}[t]{1\linewidth}
\caption{topological classification of 2D nonsymmorphic crystalline superconductors(glide plane symmetry here) with TRS. The class of Hamiltonian and its corresponding topological classification are listed in one block. $NoGS$ in the first column represents a system has no glide plane symmetry. `-' stands for no topological classification.}
\hspace{-2cm}
\label{class2}
\begin{tabular}
[c]{|c|c|c|c|c|c|c|}\hline
&\multicolumn{3}{|c|}{spinless} &\multicolumn{3}{|c|}{spin-$\frac{1}{2}$}\\\hline
NoGS(2D)&\multicolumn{3}{|c|}{BDI, -} & \multicolumn{3}{|c|}{DIII, $\mathbb{Z}_2$} \\\hline
& $\bold{k} \cdot \bold{\tau} = 0$ & $\bold{k} \cdot \bold{\tau} = \frac{\pi}{2}$& $\bold{k} \cdot \bold{\tau} \neq 0, \frac{\pi}{2}$ & $\bold{k} \cdot \bold{\tau} = 0 $ & $\bold{k} \cdot \bold{\tau} = \frac{\pi}{2}$ &$\bold{k} \cdot \bold{\tau} \neq 0, \frac{\pi}{2}$\\\hline
$G_-$ & AI, -                &D, $\mathbb{Z}_2$  & A, -&  D, $\mathbb{Z}_2$ & AII, - &A, - \\ \hline
& $\bold{k} \cdot \bold{\tau} = 0$ & $\bold{k} \cdot \bold{\tau} = \frac{\pi}{2}$ & $\bold{k} \cdot \bold{\tau} \neq 0, \frac{\pi}{2}$ & $\bold{k} \cdot \bold{\tau} = 0 $ & $\bold{k} \cdot \bold{\tau} = \frac{\pi}{2}$& $\bold{k} \cdot \bold{\tau} \neq 0, \frac{\pi}{2}$\\\hline
$G_+$ & BDI, $\mathbb{Z}\oplus \mathbb{Z}$   &AIII, $\mathbb{Z}$ &AIII, $\mathbb{Z}\oplus \mathbb{Z}$               & AIII, $\mathbb{Z}$ & DIII, $\mathbb{Z}_2\oplus \mathbb{Z}_2$ &AIII,    $\mathbb{Z}\oplus \mathbb{Z}$   \\\hline
\end{tabular}
  \end{minipage}
\end{table}
\end{center}


\subsubsection{class BDI: $G_+$ for spinless superconductors}
The TR operator is $T = \left ( \begin{array}{cc} \Theta_e & 0 \\ 0&\Theta^{\dag T}_e \end{array} \right)$ with $\Theta_e = K$ for spinless systems. One can easily check that $T G_+(\bold{k}) T^{-1} = G_+(-\bold{k})$ and also $C G_+(\bold{k}) C^{-1} = G_+(-\bold{k})$. For a state $\psi_{\bold{k}}$ with glide parity $\delta e^{i \bold{k} \cdot \bold{\tau}}$ ($\delta = \pm 1$), its PHS partner $\tilde{\psi}_{-\bold{k}}$ and TRS partner $\bar{\psi}_{-\bold{k}}$ have glide parities $\delta^* e^{-i \bold{k} \cdot \bold{\tau}}$ and $\delta^* e^{-i \bold{k} \cdot \bold{\tau}}$, respectively. Thus, at $\bold{k} \cdot \bold{\tau} = 0$, PHS and TRS coexist in each glide parity subspace. Therefore, the symmetry class in each glide parity subspace is BDI with $\mathbb{Z}$ classification in 1D. The topological invariant for the full BdG Hamiltonian is $\mathbb{Z}\oplus \mathbb{Z}$ at $\bold{k} \cdot \bold{\tau} = 0$, since the two glide parity subspaces at momentum $\bold{k} \cdot \bold{\tau} = 0$ are independent and there is no symmetry operation that can couple them. For the momentum ${\bf k\cdot\tau}=C$ with $0<C<\pi/a$, although both TRS and PHS do not exist in each glide parity subspace, the chiral symmetry, defined as the combination of TRS and PHS ($\Pi=TC$), is preserved in each glide parity subspace. Thus, for any other momentum lines, the symmetry class in each glide parity subspace is AIII with a $\mathbb{Z}$ classification in 1D. The topological invariant for the full BdG Hamiltonian is $\mathbb{Z}\oplus \mathbb{Z}$ at any momentum line $0<\bold{k} \cdot \bold{\tau}<\pi/2$, since the two glide parity subspaces are independent. The TRS or PHS can relate a state at the momentum $\bold{k} \cdot \bold{\tau}=-C$ to a state with the same glide parity at the momentum $\bold{k} \cdot \bold{\tau}=C$, which suggests that topological classifications in the momentum regime $-\pi/2<\bold{k} \cdot \bold{\tau}<0$ is directly determined by that in the momentum regime $0<\bold{k} \cdot \bold{\tau}<\pi/2$. At the momentum line $\bold{k} \cdot \bold{\tau}=\pi/2$, chiral symmetry still exists (AIII class), and TRS or PHS relate opposite glide parity subspaces. As a result, the classification should be determined only by one integer $\mathbb{Z}$.

\subsubsection{class BDI: $G_-$ for spinless superconductors}
In this case, we have $T G_-(\bold{k}) T^{-1} = G_-(-\bold{k})$ and $C G_-(\bold{k}) C^{-1} = -G_-(-\bold{k})$. For a state $\psi_{\bold{k}}$ with glide parity $\delta e^{i \bold{k} \cdot \bold{\tau}}$ ($\delta = \pm 1$), its PHS partner $\tilde{\psi}_{-\bold{k}}$ and TRS partner $\bar{\psi}_{-\bold{k}}$ have glide parities $-\delta^* e^{-i \bold{k} \cdot \bold{\tau}}$ and $\delta^* e^{-i \bold{k} \cdot \bold{\tau}}$, respectively. Thus, at $\bold{k} \cdot \bold{\tau} = 0$, PHS interchanges two states in opposite glide parity subspaces but TRS exists in each glide parity subspace. Therefore, the symmetry class in each glide parity subspace is AI with no topological classification in 1D. At $\bold{k} \cdot \bold{\tau} = \frac{\pi}{2}$, the situation is exactly opposite and the corresponding symmetry class in each glide parity subspace is D with a $\mathbb{Z}_2$ topological invariant. The topological invariant for the full BdG Hamiltonian is $\mathbb{Z}_2$ at $\bold{k} \cdot \bold{\tau} = \frac{\pi}{2}$, since the two glide parity subspaces at momentum $\bold{k} \cdot \bold{\tau} = \frac{\pi}{2}$ are related by TRS. For momentum lines $\bold{k} \cdot \bold{\tau} \neq 0, \frac{\pi}{2}$, the symmetry class is A, leading to no topological invariant in 1D.

\subsubsection{class DIII: $G_+$ for spin-$\frac{1}{2}$ superconductors}
In this case, the TR operator is $T = \left ( \begin{array}{cc} \Theta_e & 0 \\ 0&\Theta^{\dag T}_e \end{array} \right)$ with $\Theta_e = i\sigma_0s_2K$ where $\sigma$ acts on different atomic sites(orbitals) and $s$ acts on the spin space. Since $\Theta_e D_{\bold{k}}(g) \Theta^{-1}_e = D_{-\bold{k}}(g)$, $T G_+(\bold{k}) T^{-1} = G_+(-\bold{k})$ and $C G_+(\bold{k}) C^{-1} = G_+(-\bold{k})$. Thus, for a state $\psi_{\bold{k}}$ with glide parity $\delta e^{i \bold{k} \cdot \bold{\tau}}$ ($\delta = \pm i$), its PHS partner $\tilde{\psi}_{-\bold{k}}$ and TRS partner $\bar{\psi}_{-\bold{k}}$ have glide parities $\delta^* e^{-i \bold{k} \cdot \bold{\tau}}$ and $\delta^* e^{-i \bold{k} \cdot \bold{\tau}}$, respectively. Thus, at $\bold{k} \cdot \bold{\tau} = \frac{\pi}{2}$, PHS, TRS and chiral symmetries coexist in each glide parity subspace, leading to symmetry class DIII with a $\mathbb{Z}_2$ classification in 1D. Because there is no symmetry relating these two glide parity subspaces, the topological invariant for the full BdG Hamiltonian is $\mathbb{Z}_2\oplus \mathbb{Z}_2$ at $\bold{k} \cdot \bold{\tau} = \frac{\pi}{2}$. The chiral symmetry $\Pi$ exists at any other momenta in each glide parity subspace, yielding symmetry class AIII with a $\mathbb{Z}$ topological classification in 1D. For the momentum line ${\bf k}\cdot{\tau}=C$ ($0<C<\pi/2$), the topological classification is $\mathbb{Z}\oplus \mathbb{Z}$, which also determine the classification at the momentum line ${\bf k}\cdot{\tau}=-C$. It should be pointed out that in this momentum regime, the difference between class BDI and DIII lies in the fact that for DIII, TRS or PHS relate the AIII topological invariants in opposite glide parity subspaces at the momentum line ${\bf k}\cdot{\tau}=C$ and ${\bf k}\cdot{\tau}=-C$ while TRS or PHS relate the AIII topological invariants in the same glide parity subspaces for BDI class. At the momentum line ${\bf k}\cdot{\tau}=0$, chiral symmetry exists (AIII class), and TRS or PHS changes glide parities. Therefore, the classification should be determined only by one integer $\mathbb{Z}$.



\subsubsection{class DIII: $G_-$ for spin-$\frac{1}{2}$ superconductors}
In this case, $T G_-(\bold{k}) T^{-1} = G_-(-\bold{k})$ and $C G_-(\bold{k}) C^{-1} = - G_-(-\bold{k})$. The PHS partner $\tilde{\psi}_{-\bold{k}}$ and TRS partner $\bar{\psi}_{-\bold{k}}$ of a state $\psi_{\bold{k}}$ have glide parities $-\delta^* e^{-i \bold{k} \cdot \bold{\tau}}$ and $\delta^* e^{-i \bold{k} \cdot \bold{\tau}}$, respectively. Thus, at $\bold{k} \cdot \bold{\tau} = 0$, TRS interchanges two glide parity subspaces but PHS exists in each glide parity subspace. The corresponding symmetry class is D with a $\mathbb{Z}_2$ classification in 1D in each glide parity subspace. The topological invariant for the full BdG Hamiltonian is $\mathbb{Z}_2$ at $\bold{k} \cdot \bold{\tau} = 0$, since the two glide parity subspaces at momentum $\bold{k} \cdot \bold{\tau} = 0$ are related by TRS. At $\bold{k} \cdot \bold{\tau} = \frac{\pi}{2}$, PHS interchanges two glide parity subspaces but TRS exists in each glide parity subspace, giving symmetry class AII without any classification in 1D. For momentum lines $\bold{k} \cdot \bold{\tau} \neq 0, \frac{\pi}{2}$, the symmetry class is A, leading to no topological invariant in 1D.

\section{Models and topological invariants for TNCSc in different symmetry classes}
The toy model for TNCSc in the D class has been discussed in the main text. In this section, we will study models, as well as the related topological invariants, for TNCSc in other symmetry classes and show physical consequence of our classification in the Table \ref{class1} and \ref{class2}.

\subsection{Class BDI and DIII in $G_-$ configuration}
 We first consider the case of $G_-$ in the symmetry classes BDI and DIII. In both case, the chiral symmetry $\Pi$ anti-commutes with the glide symmetry, $\{\Pi,G_-({\bf k})\}$. This exactly corresponds to the situation discussed in Ref. \onlinecite{shiozaki2015}, in which the model Hamiltonian belongs to AIII class with glide symmetry and the correspond topological classification is $\mathcal{Z}_2$. For the AIII class, Dirac type of edge modes are unpinned\cite{shiozaki2015}. For our case of BDI and DIII, we can also consider the model used in Shiozaki's paper\cite{shiozaki2015} by properly imposing additional TRS and PHS. The corresponding topological classification is the same ($\mathcal{Z}_2$), but the PHS and TRS yield the gapless point of edge modes pinned at the momentum $\bold{k} \cdot \bold{\tau} = \pi/2$ for the BDI class and $\bold{k} \cdot \bold{\tau} = 0$ for the DIII class. Topological invariants can also be defined in a similar manner as in Ref. \cite{shiozaki2015}.



\subsection{Class BDI in $G_+$ configuration}
In this subsection, we will study a 2D model with $G_+$ symmetry in BDI class. The model reads \begin{eqnarray}
\nonumber H_{BDI}(\bold{k}) = &&(\epsilon(\bold{k}) - \mu)\tau_3\sigma_0 + t_3cos^2(\frac{k_xa}{2}) \tau_3\sigma_1 +  t_3cos(\frac{k_xa}{2})sin(\frac{k_xa}{2})\tau_3\sigma_2 \\
&&- \Delta_0 sin(k_ya) \tau_2\sigma_0 -\Delta_1 sin(\frac{k_xa}{2})cos(\frac{k_xa}{2})\tau_2\sigma_1 -\Delta_1 sin^2(\frac{k_xa}{2})\tau_2\sigma_2,\label{appendix_HBDI}
\end{eqnarray}
where $\epsilon(\bold{k}) = m_0 + t_1cos(k_xa) + t_2cos(k_ya)$, and $\tau_i$ and $\sigma_i$ are Pauli matrices acting in the Nambu and sublattice space, respectively. The TRS and PHS operators read $T_{BDI} = \tau_0 \sigma_0 K$ and $C_{BDI} = \tau_1 \sigma_0 K$. One can easily check that $T_{BDI} H_{BDI}(\bold{k}) T_{BDI}^{-1} = H_{BDI}(-\bold{k})$ and  $C_{BDI} H_{BDI}(\bold{k}) C_{BDI}^{-1} = -H_{BDI}(-\bold{k})$. The chiral symmetry operator is $\Pi_{BDI} = T_{BDI}\times C_{BDI} = \tau_1\sigma_0$ and $\Pi_{BDI} H_{BDI}(\bold{k}) \Pi_{BDI}^{-1} = -H_{BDI}(\bold{k})$. The glide plane symmetry operator $G_+$ is the same as that in the main text, which reads $G_{+,BDI}(\bold{k}) = \left( \begin{array} {cc} D_{\bold{k}}(g)&0\\ 0& D^*_{-\bold{k}}(g) \\ \end{array} \right)
$ with $D_{\bold{k}}(g) = e^{i\frac{k_xa}{2}}(cos(\frac{k_xa}{2}) \sigma_1 + sin(\frac{k_xa}{2}) \sigma_2)$.

As we have demonstrated previously, the chiral symmetry exists for all momenta $k_x$ in each glide parity subspace, leading to BDI class at $\bold{k\cdot\tau} = 0$ and AIII class otherwise. The topological invariant for both BDI class and AIII class in 1D is described by a winding number\cite{schnyder2008,schnyder2011,tewari2012}
\begin{eqnarray}
\nu_{\zeta_\pm,k_x} = \frac{1}{2\pi i} \oint_{\textit{L}} dk_y Tr[Q^{-1}_{\zeta_\pm}(\bold{k})\nabla_{k_y}Q_{\zeta_\pm}(\bold{k})]
\label{winding}
\end{eqnarray}
for the subspace with glide parity $\zeta_\pm = \pm e^{\frac{ik_xa}{2}}$. The integral is applied along a closed loop $L$ at momentum $k_x$ in the Brillouin zone. Due to chiral symmetry existing in one glide parity subspace, we can always find a unitary matrix $U$ to transform the Hamiltonian into an off-block-diagonal form\cite{schnyder2008,schnyder2011}, $UH_{BdG,\zeta_\pm}U^\dag = \left( \begin{array} {cc} 0&Q_{\zeta_{\pm}}(\bold{k})\\Q_{\zeta_\pm}^\dag(\bold{k})&0\\ \end{array} \right)$, in which $Q_{\zeta_\pm}$ is the off-diagonal block. For our model, we can consider the integral loop $L$ along the $k_y$ direction and regard $k_x$ as a parameter. Therefore, $\nu_{\zeta_\pm}$ is a function of $k_x$ and Eq. (\ref{winding}) can be further simplified as $\nu_{\zeta_\pm,k_x} = \frac{1}{2\pi i} \oint_{\textit{L}} d[ln(det Q_{\zeta_\pm}(k_x,k_y))]$, which suggests that the winding number is related to how many loops $det [Q_{\zeta_\pm}(k_x,k_y)]$ evolves around the origin in the complex plane as $k_y$ changes from $-\pi$ to $\pi$. Since TRS relates two states in the same glide subspace, one can easily prove the relation $\nu_{\zeta_\pm,k_x} = \nu_{\zeta_\pm,-k_x}$.

To verify the Z topological invariant for BDI class in $G_+$ configuration, we consider a semi-infinite system for the Hamiltonian (\ref{appendix_HBDI}) with an open boundary along the y direction and apply the iterative Green function method\cite{sancho1984} to calculate local density of states (LDOS) at the boundary, as illustrated in Fig. \ref{figBDI}(a), with the parameters listed in Table \ref{par3}. Strikingly, we find two-fold degenerate zero energy flat bands for all momenta $k_x$ from our calculations.

\begin{figure}[tb]
	\includegraphics[width = 0.9\columnwidth,angle=0]{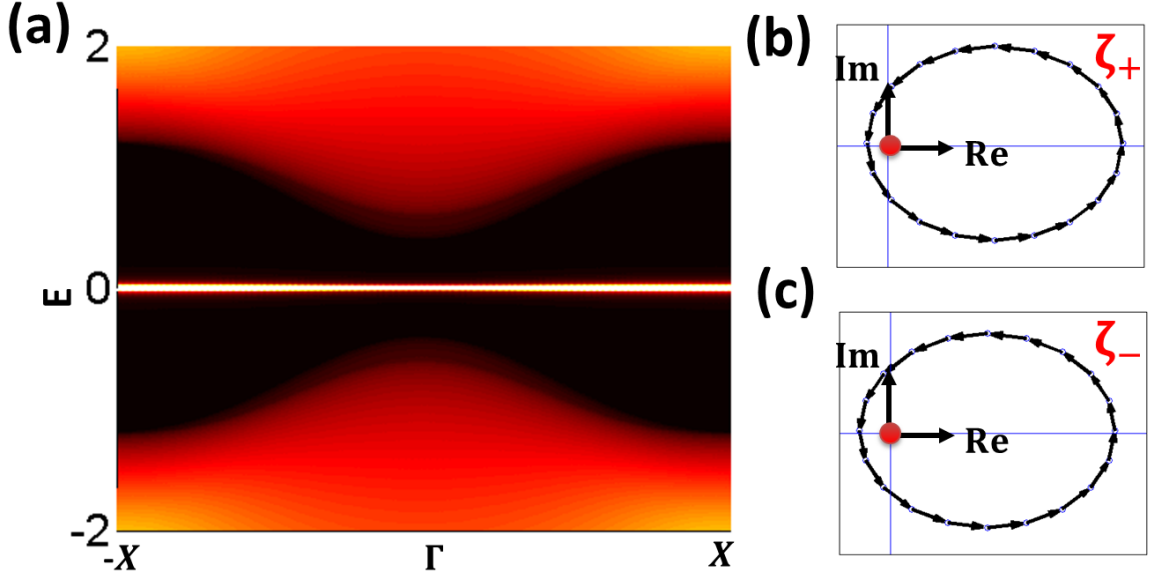}
  \caption{  (Color online). (a) Edge DOS with zero energy flat bands appearing for BDI class in $G_+$ configuration. (b) Illustration of winding number in glide parity $\zeta_+ = e^{\frac{ik_xa}{2}}$ subspace. (c) Illustration of winding number in glide parity $\zeta_- = -e^{\frac{ik_xa}{2}}$ subspace. The red dot denotes the origin in the complex plane.}
    \label{figBDI}
\end{figure}

\begin{center}
\begin{table}[htb]
  \centering
  \begin{minipage}[t]{1.\linewidth}
	  \caption{Parameters for $G_{+}$ configuration in class BDI.}
\hspace{-1cm}
\begin{tabular}
[c]{cccccccc}\hline\hline
      &$m_0$& $t_1$& $t_2$& $t_3$ & $\mu$ & $\Delta_0$ & $\Delta_1$\\\hline
BDI: $G_+$ & 1.5& 0.5& 2.5& 0.1& 0& 2 & 0.5\\\hline
\label{par3}
\end{tabular}
  \end{minipage}
\end{table}
\end{center}

In order to demonstrate topological property of these flat bands, we perform an analytical calculation for topological invariant. The BdG Hamiltonian (\ref{appendix_HBDI}) can be written in a block-diagonal form $H_{BDI} = diag[H_{\zeta_+ = e^{i\frac{k_xa}{2}}},H_{\zeta_- = -e^{i\frac{k_xa}{2}}}]$, where $H_{\zeta_\pm} = (\epsilon(\bold{k})\pm t_3 cos(\frac{k_xa}{2}))\tau_3 - \Delta_0 sin(k_ya)\tau_2 - \Delta_1 sin(\frac{k_xa}{2})\tau_2$. Each block has a specific glide parity $\zeta_\pm$. By performing a further basis transformation, we obtain the off-diagonal block $Q_{\zeta_\pm}$ for $H_{\zeta_\pm}$, which is $Q_{\zeta_\pm} = \epsilon(\bold{k}) \pm t_3 cos(\frac{k_xa}{2}) - i \Delta_0 sin(k_ya) -i \Delta_1sin(\frac{k_xa}{2})$. The winding number $\nu_{\zeta_\pm}$ for $Q_{\zeta_\pm}$ is mainly determined by $Delta_0$ and $t_2$ in this case and for our choice of parameters, $\nu_{\zeta_\pm}(k_x) = 1$ for all momenta $k_x$. We plot the evolution of $det [Q_{\zeta_\pm}]$ for $k_xa = 0.2$, as shown in Fig. \ref{figBDI}(b) and (c), which confirm the non-trivial winding number for our model. Due to non-zero winding number $\nu_{\zeta_\pm}(k_x) = 1$, zero energy modes (Majorana modes) emerge for any momentum $k_x$, leading to zero energy flat bands. In addition, the relation $\nu_{\zeta_\pm,k_x} = \nu_{\zeta_\pm,-k_x}$ due to TRS suggests that Majorana flat bands should be symmetric around ${\bf k}=0$ in each glide parity subspace.

\subsection{Class DIII in $G_+$ configuration}
For NTCSc in class DIII, we consider the same lattice structure as that in the main text. The glide plane symmetry operator for the normal state reads $D_{\bold{k}}(g) = i e^{i\frac{k_xa}{2}}(cos(\frac{k_xa}{2}) \sigma_1 s_3 + sin(\frac{k_xa}{2}) \sigma_2 s_3)$ on the basis $c_{\bold{k}} = (|A,\bold{k}, \uparrow\rangle, |A,\bold{k}, \downarrow\rangle, |B,\bold{k},\uparrow\rangle, |B,\bold{k},\downarrow\rangle)^T$ where $\sigma_i$ and $s_i$ are the Pauli matrices acting in the sublattice and spin space. The TRS operator is $\Theta = i\sigma_0s_2K$. The Hamiltonian, on such a basis $c_{\bold{k}}$, is expressed as $h_e(\bold{k}) = \epsilon(\bold{k}) \sigma_0s_0 + t_4sin(k_xa)\sigma_3s_1 - t_5 sin(k_ya) \sigma_3 s_2 + t_3sin(\frac{(k_x-\phi)a}{2}) (cos(\frac{k_xa}{2})\sigma_1+ sin(\frac{k_xa}{2})\sigma_2)(\frac{s_3+s_0}{2}) + t_3 sin(\frac{(k_x+\phi)a}{2})(cos(\frac{k_xa}{2})\sigma_1+ sin(\frac{k_xa}{2})\sigma_2)(\frac{s_3-s_0}{2})$, where $\epsilon(\bold{k}) = m_0 + t_1cos(k_xa) + t_2cos(k_ya)$. One can easily check that this Hamiltonian satisfies that $D^{-1}_{\bold{k}}(g) h_{e}(\bold{k}) D_{\bold{k}}(g) = h_e(\bold{k})$ and $\Theta^{-1} h_e(\bold{k}) \Theta = h_e(-\bold{k})$.

On the basis $\Psi(\bold{k}) = (c_{\bold{k}}, c^{\dag T}_{-\bold{k}})^T$ in the Nambu space, we can construct the BdG Hamiltonian in class DIII, which reads
\begin{eqnarray}
\nonumber H_{DIII} = && \epsilon(\bold{k}) \tau_3\sigma_0s_0 + t_4sin(k_xa)\tau_0\sigma_3s_1 - t_5 sin(k_ya) \tau_3 \sigma_3 s_2 \\
\nonumber && + t_3sin(\frac{(k_x-\phi)a}{2}) (\frac{\tau_0+\tau_3}{2}) (cos(\frac{k_xa}{2})\sigma_1+ sin(\frac{k_xa}{2})\sigma_2)(\frac{s_3+s_0}{2})\\
\nonumber && + t_3sin(\frac{(k_x-\phi)a}{2}) (\frac{\tau_0-\tau_3}{2}) (cos(\frac{k_xa}{2})\sigma_1+ sin(\frac{k_xa}{2})\sigma_2)(\frac{s_3-s_0}{2}) \\
\nonumber && + t_3 sin(\frac{(k_x+\phi)a}{2})(\frac{\tau_0-\tau_3}{2})(cos(\frac{k_xa}{2})\sigma_1+ sin(\frac{k_xa}{2})\sigma_2)(\frac{s_3+s_0}{2}) \\
\nonumber &&+  t_3 sin(\frac{(k_x+\phi)a}{2})(\frac{\tau_0+\tau_3}{2})(cos(\frac{k_xa}{2})\sigma_1+ sin(\frac{k_xa}{2})\sigma_2)(\frac{s_3-s_0}{2})\\
&& + \Delta_0 sin(k_ya) \tau_1\sigma_0s_1
\end{eqnarray}
where $\tau_i$, $\sigma_i$ and $s_i$ are Pauli matrices acting on the Nambu space, sublattice space and spin space, respectively. The glide plane symmmetry, TRS, PHS and chiral symmetry operators read $G_{+,DIII}(\bold{k}) = \left( \begin{array} {cc} D_{\bold{k}}(g)&0\\ 0& D^*_{-\bold{k}}(g) \\ \end{array} \right)$, $T_{DIII} = i\tau_0\sigma_0s_2 K$, $C_{DIII} = \tau_1\sigma_0s_0 K$ and $\Pi_{DIII} = \tau_1\sigma_0s_2$(Note that we ignore the `i' in the chiral symmetry operator, which is not essential).

There is chiral symmetry in each glide parity subspace for any momenta, and thus, similar to the BDI case, we can also apply the 1D winding number (E1. \ref{winding}) as the topological invariant to this model. One can find a unitary matrix V to transform the Hamiltonian into an off-block-diagonal form\cite{schnyder2011}, $VH_{BdG}V^\dag = \left( \begin{array} {cc} 0&q(\bold{k})\\q^\dag(\bold{k})&0\\ \end{array} \right)$ with $q^T(-\bold{k}) = -q(\bold{k})$ and the corresponding winding number is defined as
\begin{eqnarray}
\nu_{\zeta_\pm,k_x} = \frac{1}{2\pi i} \oint_{\textit{L}} dk_y Tr[q^{-1}_{\zeta_\pm}(\bold{k})\nabla_{k_y}q_{\zeta_\pm}(\bold{k})]
\label{winding2}
\end{eqnarray}
where $q_{\zeta_\pm}$ is the block of the matrix $q$ in the glide parity $\zeta_{\pm}$ subspace. For DIII class, TRS relate states in opposite glide parity subspaces, and thus one can show $\nu_{\zeta_+,k_x} = - \nu_{\zeta_-,-k_x}$. However, for momentum line $\bold{k \cdot \tau} = \frac{\pi}{2}$, we have $\nu_{\zeta_+,k_x=\pi/a} = - \nu_{\zeta_-,k_x=-\pi/a}=- \nu_{\zeta_+,k_x=\pi/a}$, indicating that $\nu_{\zeta_{\pm},k_x=\pi/a}=0$. This conclusion is consistent with the fact that the Hamiltonian in each glide parity subspace belongs to symmetry class DIII with TRS. However, due to the anti-symmetric nature of the matrix $q$, an independent $\mathcal{Z}_2$ topological invariant\cite{schnyder2011} can be defined as
\begin{eqnarray}
\textit{W}_{\zeta_\pm,k_x = \frac{\pi}{a}} = \Pi_{\mathcal{K}} \left\{Pf[q^T_{\zeta_\pm}(\mathcal{K})]/\sqrt{det[q_{\zeta_\pm}(\mathcal{K})]}\right\}
\label{Wz2},
\end{eqnarray}
where $\zeta_\pm$ denotes glide parity $\zeta_\pm = \pm ie^{\frac{ik_xa}{2}}$ subspace,
$\textit{L}$ is a loop with momentum $\frac{\pi}{a}$ in the BZ, $\mathcal{K}$ denotes time reversal invariant momenta $(k_x,k_y) =(\frac{\pi}{a}, 0)$, $(\frac{\pi}{a}, \frac{\pi}{a})$ and $Pf$ denotes the Pfaffian. $\textit{W}_{\zeta_\pm,\frac{\pi}{a}} = \pm 1$ denotes trivial and nontrivial topological phases.

The local DOS at the edge for a semi-infinite system is shown in Fig. \ref{figDIII}(a) with the parameters in Table \ref{par4}. Interestingly, we find two types of edge modes in the energy dispersion at the boundary. There are zero energy flat bands, similar to the case of BDI class, around $\Gamma$ ($k_x=0$), and a Dirac type of energy dispersion with gapless point at $X$ ($k_x=\pi/a$). To understand edge modes in Fig. \ref{figDIII}(a), we study topological invariants in each glide parity subspace. The Hamiltonian in each glide parity subspace is written as $H_{DIII, \zeta_\pm = \pm i e^{i\frac{k_xa}{2}}} = \epsilon(\bold{k})\tau_3\sigma_0 \pm  t_3sin(\frac{(k_x-\phi)a)}{2})(\frac{\tau_3+\tau_0}{2})(\frac{\sigma_0+\sigma_3}{2}) \pm  t_3sin(\frac{(k_x-\phi)a)}{2})(\frac{\tau_3-\tau_0}{2})(\frac{\sigma_0-\sigma_3}{2}) \pm  t_3sin(\frac{(k_x+\phi)a)}{2})(\frac{\tau_3+\tau_0}{2})(\frac{\sigma_0-\sigma_3}{2}) \pm
t_3sin(\frac{(k_x+\phi)a)}{2})(\frac{\tau_3-\tau_0}{2})(\frac{\sigma_0+\sigma_3}{2}) + t_4 sin(k_xa) \tau_0\sigma_1 -t_5 sin(k_ya)\tau_z\sigma_2 + \Delta_0 sin(k_ya) \tau_1\sigma_1$. The local DOS in each glide parity $\zeta_\pm$ subspace are shown in Fig. \ref{figDIII}(c) and (d).
We calculate the winding number $\nu_{k_x}$ as a function momentum $k_x$ for $H_{DIII, \zeta_\pm}$, as shown in Fig. \ref{figDIII}(b). The zero energy flat bands exist in the momentum regime of $k_x$ where $\nu_{\zeta_\pm,k_x} = \pm 1$ and disappear in the momentum regime with $\nu_{\zeta_\pm,k_x} = 0$. Thus, the flat bands originate from the winding number $\nu_{\zeta_\pm,k_x}$, similar to the case of class BDI. At the momentum where the winding number $\nu_{\zeta_\pm,k_x}$ changes between 0 and 1, the bulk superconducting gap closes. Thus, our model corresponds to a nodal superconductor. At the momentum $k_x=\pi/a$, although $\nu_{\zeta_\pm,k_x} = 0$, we find $\textit{W}_{\zeta_\pm}=1$, giving rise to Dirac type of edge modes. This confirms two independent topological invariants in our case.



\begin{figure}[tb]
	\includegraphics[width = 0.9\columnwidth,angle=0]{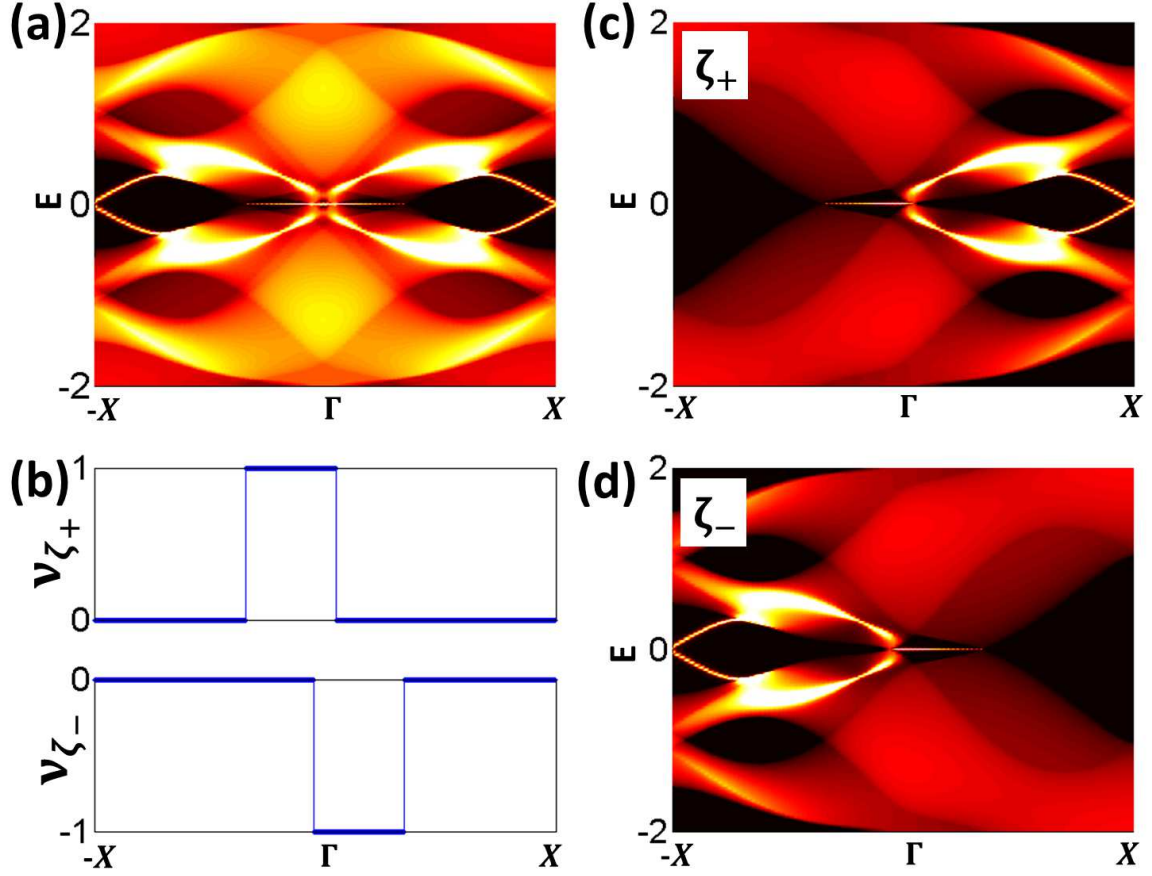}
  \caption{  (Color online). (a) Edge DOS with both zero energy flat bands and helical edge modes for DIII class in $G_+$ configuration. (b) Winding number $\nu_{\zeta_\pm}$ in glide parity $\zeta_\pm =\pm ie^{\frac{ik_xa}{2}}$ subspace as a function of $k_x$. (c) Edge DOS in glide parity subspace $\zeta_+ = ie^{\frac{ik_xa}{2}}$. (d) Edge DOS in glide parity  $\zeta_- = -ie^{\frac{ik_xa}{2}}$ subspace.}
    \label{figDIII}
\end{figure}

\begin{center}
\begin{table}[htb]
  \centering
  \begin{minipage}[t]{1.\linewidth}
	  \caption{Parameters for $G_{+}$ configuration in class DIII.}
\hspace{-1cm}
\begin{tabular}
[c]{cccccccccc}\hline\hline
      &$m_0$& $t_1$& $t_2$& $t_3$ & $t_4$ &$t_5$&$\mu$ & $\Delta_0$ &$\phi$\\\hline
DIII: $G_+$ &-1 & 0.1& 1& 1& 0.5& 0.5&0 & 1 & 0.1$\pi$\\\hline
\label{par4}
\end{tabular}
  \end{minipage}
\end{table}
\end{center}

\subsection{Chern number in Class D}
Finally, we will show that the conventional Chern number can also be defined in the whole Brillouin zone for nonsymmorphic superconductors in the D class, which just correspond to chiral topological superconductors. However, in this case, glide plane symmetry does not play a key role in the sense that Chern number is still well defined even if glide symmetry is broken.

The model Hamiltonian of the normal state is the same as the model listed in the main text. The gap function reads $\Delta(\bold{k}) = \Delta_0 (sin(k_x) + isin(k_y))$, which belongs to $G_+$ configuration. We perform a calculation of edge density of state(DOS) to illustrate the chiral edge mode explicitly by using iterative Green function method\cite{sancho1984}, as shown in Fig. \ref{figD}. Here the open boundary is applied in the y direction and we consider the upper edge that parallel to x direction. The parameters we use are listed in Table \ref{par2}. From Fig. \ref{figD}, we find one edge mode with positive velocity (a right mover), which is consistent with our calculation of Chern number $n = 1$.

\begin{figure}[tb]
	\includegraphics[width = 0.7\columnwidth,angle=0]{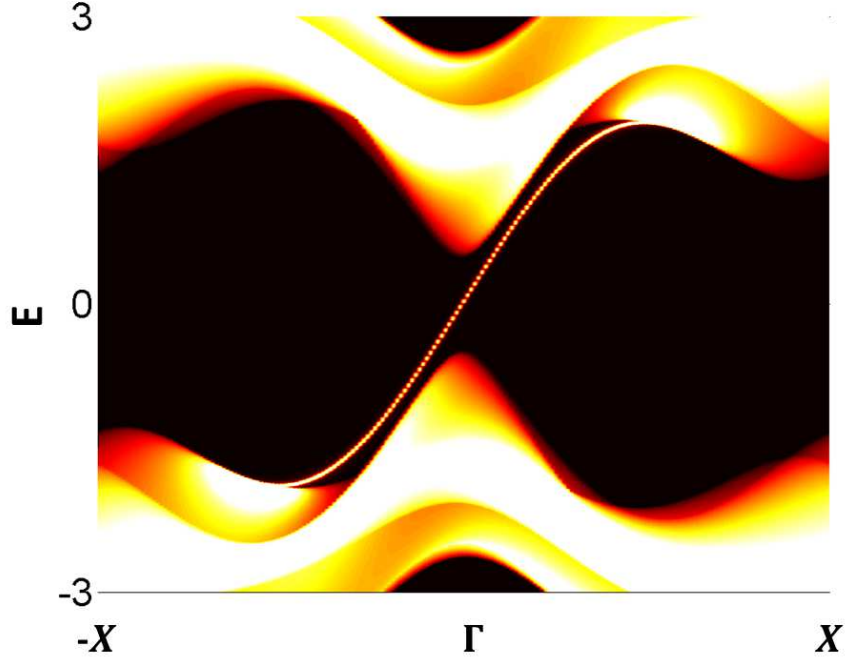}
  \caption{  (Color online). Edge DOS with a chiral edge state appearing along the x direction.
  }
    \label{figD}
\end{figure}

\begin{center}
\begin{table}[htb]
  \centering
  \begin{minipage}[t]{1.\linewidth}
	  \caption{Parameters for $G_{+}$ configuration in class D.}
\hspace{-1cm}
\begin{tabular}
[c]{cccccccc}\hline\hline
      &$m_0$& $t_1$& $t_2$& $t_3$ & $\mu$ & $\phi$ & $\Delta_0$\\\hline
Chern: $G_+$ & 1.5& -1& -1& -1& 0& 0.1$\pi$& 2\\\hline
\label{par2}
\end{tabular}
  \end{minipage}
\end{table}
\end{center}

\end{widetext}
\end{appendix}

\bibliography{NonSimmTSC}

\end{document}